\documentclass[11pt]{article}
\usepackage[margin=1in]{geometry}
\usepackage[T1]{fontenc}
\usepackage[utf8]{inputenc}
\usepackage{lmodern,microtype}
\usepackage{amsmath,amssymb,amsthm,mathtools}
\usepackage{enumitem,booktabs,array,needspace,etoolbox,float}
\usepackage{tikz}
\usetikzlibrary{arrows.meta,positioning,calc}
\usepackage{xurl}
\usepackage{hyperref}
\newcommand{\doi}[1]{\href{https://doi.org/#1}{\nolinkurl{#1}}}
\definecolor{linkblue}{rgb}{0,0.25,0.65}
\hypersetup{colorlinks=true,linkcolor=linkblue,citecolor=linkblue,
  urlcolor=linkblue,pdftitle={Product-Linear POSIX Parsing}}
\allowdisplaybreaks

\if0

\fi

\newtheorem{theorem}{Theorem}[section]
\newtheorem{lemma}[theorem]{Lemma}
\newtheorem{proposition}[theorem]{Proposition}
\newtheorem{corollary}[theorem]{Corollary}

\theoremstyle{definition}
\newtheorem{definition}[theorem]{Definition}
\AtBeginEnvironment{definition}{\Needspace{8\baselineskip}\pushQED{\qed}}
\AtEndEnvironment{definition}{\popQED}

\newtheorem{example}[theorem]{Example}
\theoremstyle{remark}
\newtheorem{remark}[theorem]{Remark}

\apptocmd{\thebibliography}{\setlength{\itemsep}{2pt plus .5pt minus .5pt}}{}{}

\newcommand{\eps}{\varepsilon}
\newcommand{\emptyre}{\varnothing}
\newcommand{\capture}[2]{\langle #2\rangle_{#1}}
\newcommand{\RepExpr}{\mathsf{Rep}}
\newcommand{\lo}{\mathrm{lo}}
\newcommand{\hi}{\mathrm{hi}}
\newcommand{\Parse}{\mathcal P}
\newcommand{\Opt}{\mathsf{Opt}}
\newcommand{\spanop}{\mathsf{span}}
\newcommand{\len}{\mathsf{len}}
\newcommand{\codeop}{\mathsf{code}}
\newcommand{\opentok}{\mathsf{open}}
\newcommand{\closetok}{\mathsf{close}}
\newcommand{\tagop}{\mathsf{tag}}

\newcommand{\lastv}{\mathsf{last}}
\newcommand{\rep}{\mathsf{rep}}
\newcommand{\capv}{\mathsf{cap}}
\newcommand{\parcap}{\mathsf{par}}

\newcommand{\acc}{\mathsf{acc}}
\newcommand{\push}{\operatorname{push}}
\newcommand{\pop}{\operatorname{pop}}
\newcommand{\firstmax}{\mathsf{firstmax}}
\newcommand{\pre}{\operatorname{pre}}
\newcommand{\tdepth}{\operatorname{depth}}
\newcommand{\parentof}{\operatorname{parent}}
\newcommand{\dist}{\operatorname{dist}}

\newcommand{\spans}{\mathsf{spans}}

\newcommand{\dom}{\mathsf{dom}}
\newcommand{\norm}{\mathsf{norm}}
\newcommand{\flattenten}{\mathsf{flat}}

\title{Product-Linear POSIX Parsing}
\author{Yuya Uezato \\ yuya.uezato@gmail.com \quad CyberAgent, Inc.}
\date{}
\if0
\makeatletter
\renewcommand{\@maketitle}{%
  \begin{center}
    {\LARGE\@title\par}
    \ifx\@author\@empty\else\vskip 1em{\large\@author\par}\fi
    \ifx\@date\@empty\else\vskip .5em{\@date\par}\fi
  \end{center}}
\makeatother
\fi

\begin{document}
\maketitle
\begin{abstract}
POSIX regular-expression matching reports capture spans determined by the parse tree that its disambiguation rules select.
Thompson's classical construction decides whether an expression $E$ matches a text $w$ in product-linear time, $O(|E|\,|w|)$.
Whether the POSIX parse can also be computed in product-linear time has remained open since the rules were standardized in 1992.
For instance, the algorithm of Okui and Suzuki takes $O(|E|^2|w|)$ time after precomputation.
We give, to our knowledge, the first algorithm that computes the POSIX parse and all capture spans in product-linear total time, $O(|E|\,|w|)$.
It uses $O(|E|+|w|)$ working space and supports full matching,
longest-prefix matching, and leftmost-longest search. The algorithm scans the text backward and keeps one preferred continuation per traversal state.
It implements POSIX selection with lists of end positions of the open subexpressions, which it compares and updates in constant time.
\end{abstract}

\section{Introduction}\label{sec:intro}

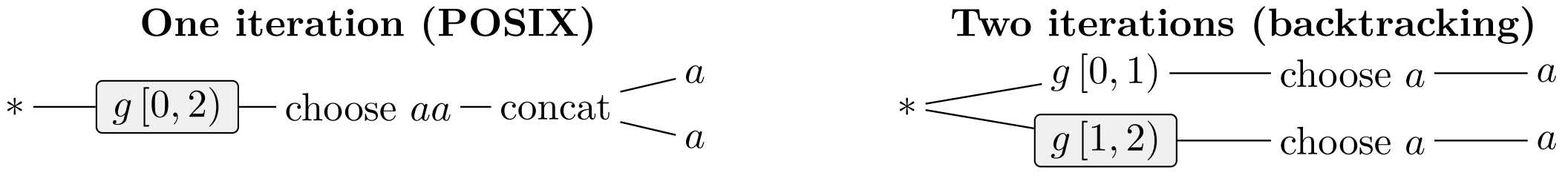
\begin{figure}[H]
\centering
\begin{tikzpicture}[x=1cm,y=1cm,font=\small,
  every node/.style={inner xsep=2pt,inner ysep=1.6pt},
  reported/.style={draw,rounded corners=1.5pt,fill=black!6,inner xsep=4pt},
  wire/.style={line width=.45pt}]
  \node[font=\small\bfseries] at (3.15,.72) {One iteration (POSIX)};
  \node (s1) at (0,0) {$*$};
  \node[reported] (g1) at (1.36,0) {$g\,[0,2)$};
  \node (r1) at (3.15,0) {choose $aa$};
  \node (c1) at (4.82,0) {concat};
  \node (a1) at (6.06,.29) {$a$};
  \node (a2) at (6.06,-.29) {$a$};
  \draw[wire] (s1)--(g1) (g1)--(r1) (r1)--(c1) (c1)--(a1) (c1)--(a2);
  \begin{scope}[xshift=7.95cm]
    \node[font=\small\bfseries] at (3,.72) {Two iterations (backtracking)};
    \node (s2) at (0,0) {$*$};
    \node (g2) at (1.77,.30) {$g\,[0,1)$};
    \node[reported] (g3) at (1.77,-.30) {$g\,[1,2)$};
    \node (l2) at (3.97,.30) {choose $a$};
    \node (l3) at (3.97,-.30) {choose $a$};
    \node (b1) at (5.70,.30) {$a$};
    \node (b2) at (5.70,-.30) {$a$};
    \draw[wire] (s2)--(g2) (s2)--(g3) (g2)--(l2) (g3)--(l3) (l2)--(b1) (l3)--(b2);
  \end{scope}
\end{tikzpicture}
\caption{The two parse trees of $(\capture{g}{a\mid aa})^*$ on \texttt{aa}.
Roots are on the left, and earlier children are above later ones. Boxes mark
the reported captures.}
\label{fig:opening-parses}
\end{figure}

Write $\capture{g}{e}$ for a capture group that records the text matched by
$e$ under the name $g$.\footnote{Juxtaposition denotes concatenation, $\mid$
alternation, and parentheses grouping. The quantifiers $*$, $+$, and $?$ allow
zero or more, one or more, and zero or one repetitions, respectively.
Positions in a word are numbered from $0$, and $[i,j)$ is the part of the word
between positions $i$ and $j$.}
How should $(\capture{g}{a\mid aa})^*$ parse \texttt{aa}?
Figure~\ref{fig:opening-parses} shows its two parse trees; both match all of
\texttt{aa}. A backtracking matcher such as Perl, PCRE, or JavaScript tries the
alternative $a$ first in every iteration. It therefore takes two iterations and
reports $g=[1,2)$, the span of the last one. POSIX instead prefers the tree
whose first iteration is longer, and reports $g=[0,2)$.

POSIX resolves the ambiguity of a regular expression by three
rules~\cite{posix2024regex,posix2024regexec}.
\begin{enumerate}[label=(\roman*),leftmargin=*,nosep]
\item Subexpressions match as long as possible, earlier ones first
      (Figure~\ref{fig:opening-parses}).
\item A group inside a repetition reports its last iteration. On
      \texttt{aaa}, $(\capture{g}{a\mid aa})^*$ takes the iterations $aa$
      and $a$, and reports $g=[2,3)$.
\item A nested group is reported only within the reported match of the group
      around it. On \texttt{ab}, $(\capture{k}{\capture{g}{a}\mid b})^*$
      reports $k=[1,2)$ and leaves $g$ unset, although $g$ matched $[0,1)$ in
      the first iteration.
\end{enumerate}

Let $m:=|E|$ be the number of nodes of the syntax tree of an expression $E$,
and $n:=|w|$ the length of a text $w$. Thompson's construction decides whether
$E$ matches $w$ in $O(mn)$ time~\cite{thompson1968}. Can the POSIX parse be
found within the same bound? The difficulty is that the POSIX choice at a node
depends on its context. Even when the matched interval is fixed, taking a
shorter branch of an alternation can let an enclosing subexpression match
more, and POSIX then prefers the shorter branch (Section~\ref{sec:examples}).
The algorithm of Okui and Suzuki~\cite{okui2011,okui2013} runs in
$O(nd(d+p))$ time after precomputing comparisons between transitions, where
$d$ and $p$ can both be $\Theta(m)$.\footnote{Here $d$ is the largest number
of literal nodes of $E$ with the same letter, and $p$ counts capture positions
and stars. The parameter $t$ in the next bound counts captures together with their
enclosing subexpressions. Section~\ref{sec:related} gives the precise
comparison and distinguishes online time from preprocessing.}
The algorithm of Borsotti and
Trofimovich~\cite{borsotti2021} runs in $O(nm^2t)$ time including
preprocessing, where $t\le m$. Both bounds are quadratic or worse in $m$ per
input character in the worst case (Table~\ref{tab:related}). Myers, Oliva, and
Guimar\~aes~\cite{myers1998reporting} find longest spans and construct
matching-path graphs in $O(mn)$ time; their general optimal-tag selection has
an additional tag-dependent factor. Those objectives differ from POSIX even
on a fixed interval (Section~\ref{sec:related}).

We allow literals, the empty word $\eps$, the empty language $\emptyre$,
concatenation, alternation, named captures, and $*$, $+$, $?$ over arbitrary
operands, including \emph{nullable} ones, which match $\eps$. We use the usual
longest-match mode, without minimal-repetition modifiers or
\texttt{REG\_MINIMAL}. Because the
POSIX choice depends on how the expression is parenthesized
(Section~\ref{sec:examples}), Section~\ref{sec:semantics} defines the POSIX
order on the syntax tree as written. POSIX also has interval expressions
$e\{k,l\}$ and back-references. We exclude both: with either feature, no
algorithm runs in $O(mn)$ time with $m$ the size of the expression as written,
unless the Orthogonal Vectors Conjecture fails.\footnote{Nested squaring
$e\{2\}$ and a single group with straight-line back-references each rule out
$O(n^{2-\varepsilon}\operatorname{poly}(m))$ time for every $\varepsilon>0$
under this conjecture~\cite[Theorems~1 and~6]{nogami2026hardness}.
Section~\ref{sec:related} discusses both features.} Full matching requires the
match to span all of $w$. Longest-prefix matching fixes the start and
maximizes the end. Leftmost-longest search takes the least feasible start and
then maximizes the end. We assume a word RAM with $\Theta(\log(m+n+2))$-bit
words, on which arithmetic, comparison, and addressing take constant time.
Auxiliary space excludes the input and the output.

\begin{theorem}[Product-linear POSIX parsing]\label{thm:main}
Full matching, longest-prefix matching, and leftmost-longest search on
$(E,w)$ take $O(m(n+1))$ deterministic time, including preprocessing. On
success the algorithm returns the POSIX parse tree and the span of every
capture group, or only the spans, using $O(m+\ell)$ auxiliary words, where
$\ell$ is the length of the match. Failure uses $O(m)$ auxiliary words.
\end{theorem}

The term $n+1$ only accounts for the empty text: for $n\ge1$ the time bound is
$O(|E|\cdot|w|)$.

\paragraph{Contributions.}
\begin{itemize}[leftmargin=*,nosep]
\item To our knowledge, Theorem~\ref{thm:main} gives the first algorithm that
      selects the POSIX parse in time linear in $m(n+1)$, the cost of
      Thompson's membership test.
\item The key ingredient is a comparison of continuations that depends only on
      the current state of a backward scan (Section~\ref{sec:continuations}),
      with keys that can be compared and updated in constant time
      (Section~\ref{sec:evaluation}).
\item We recover the parse in $O(m+\ell)$ auxiliary words by contracting the
      intervals matched by a chosen subexpression (Section~\ref{sec:outputs}).
      For $\ell>0$ we also encode it with at most $2\ell-1$ vertices; an empty
      match has one compact record. The parse tree itself can have
      $\Theta(mn)$ vertices even when it is unique
      (Example~\ref{ex:output-lower-bound}).
\item We report the capture vectors at all start positions in $O(m(n+1))$
      total time (Section~\ref{sec:further}).
\end{itemize}

\paragraph{Key ideas.}
We view a parse as a left-to-right depth-first traversal that enters nodes,
reads letters, and returns from nodes. Section~\ref{sec:continuations} shows
that the POSIX order among the continuations from a traversal state depends
only on the state. A backward scan can therefore store one preferred
continuation per state. At each choice, it compares the lists of end positions
of the open nodes, using a fixed local order to break ties. These lists can be
long, but each scan step appends only its current input position, which is
smaller than the positions introduced at previous steps. A trie rebuilt once
per position then compares and updates the lists in constant time
(Section~\ref{sec:evaluation}). Finally, to output the parse without storing a
choice for every state, we follow Bille and
G{\o}rtz~\cite{bille2019,bille2022}: we split the expression at a subtree,
contract the intervals that the subtree matches, and recurse
(Section~\ref{sec:outputs}).

Like our algorithm, two-pass parsers compute recognition information before
reconstructing a parse; the idea goes back to Kearns~\cite{kearns1991}.
Frisch and Cardelli~\cite{frischcardelli2004} compute suffix feasibility
backward and then reconstruct a greedy parse forward in $O(mn)$ time. Sulzmann
and Lu~\cite[Section~4.4]{sulzmann2010} also use backward processing to avoid
carrying partial-parse histories. For POSIX, feasible choices must additionally
be ranked by their continuations: Section~\ref{sec:examples} gives an example
in which both branches remain feasible while the preferred branch changes with
the input.

\paragraph{Organization.}
Section~\ref{sec:examples} illustrates which parse POSIX selects, and
Section~\ref{sec:overview} traces our algorithm on one expression.
Section~\ref{sec:semantics} defines the candidate parses, their order, the
queries, and the capture vector. Section~\ref{sec:continuations} reduces the
choice to a comparison of keys at traversal states, and
Section~\ref{sec:evaluation} processes one input position in $O(m)$ time.
Section~\ref{sec:outputs} reconstructs the parse and its captures in linear
space, Section~\ref{sec:further} handles several starts, and
Section~\ref{sec:related} discusses related work. The appendices summarize the
notation and contain the remaining proofs.

\section{Examples: Which parse does POSIX select?}\label{sec:examples}

Leftmost-longest search chooses the earliest feasible start, then the latest
feasible end at that start~\cite[Chapter~9]{posix2024regex}. For
$E=(\texttt{=}\mid\texttt{==})$, the search returns the first \texttt{==} in
\texttt{x==y=z}, but only the single \texttt{=} at position~$1$ in
\texttt{x=y==z}: an earlier start beats a longer match. What if several parses
cover the same interval? The next two examples require a full match.

\paragraph{The reported capture does not identify the parse.}
Match $(\capture{g}{a\mid aa})^*$ against \texttt{aaa}. It has exactly three
parses, and each is selected by a different rule (Table~\ref{tab:three-parses}).
\begin{table}[H]
\centering
\begin{tabular}{@{}lll@{}}
\toprule
Selection rule & Iterations & Reported final $g$ \\
\midrule
First feasible branch in every iteration (backtracking) & $a\;/\;a\;/\;a$ & $[2,3)$: \texttt{a} \\
Maximize the length of the final capture & $a\;/\;aa$ & $[1,3)$: \texttt{aa} \\
POSIX (Definition~\ref{sem:def-recursive-order}) & $aa\;/\;a$ & $[2,3)$: \texttt{a} \\
\bottomrule
\end{tabular}
\caption{The three parses of $(\capture{g}{a\mid aa})^*$ on \texttt{aaa}.
Slashes separate successive iterations, and the middle row uses a hypothetical
rule. In each row the reported span belongs to the final iteration of the
selected parse.}
\label{tab:three-parses}
\end{table}
POSIX compares iterations from left to right, so it selects the third row, the
only one whose first iteration has two letters. The first and third rows report
the same final capture, $[2,3)$. Thus the reported capture vector does not
determine the selected parse: two different parses can have the same vector.

\paragraph{A shorter choice can win.}\label{para:context-priority}
Let $X:=aa\mid a$, $Y:=(aa)?$, and $Z:=a?$. From here on, parentheses fix the
syntax tree. On \texttt{aaa} there are just two full decompositions:
\[
\begin{array}{@{}ccc@{\qquad}cc@{}}
 X & Y & Z & \text{end of }XY & \text{end of }X \\
 \hline
 aa & \eps & a & 2 & 2 \\
 a & aa & \eps & 3 & 1
\end{array}
\]
On the tree $(XY)Z$, POSIX compares the left child $XY$ first. Giving $X$ only
one letter lets $XY$ consume all three letters, so the second row wins. On the
tree $X(YZ)$, the left child is $X$ itself, and the first row wins. To observe
the difference, mark the short branch by writing $X:=aa\mid\capture{g}{a}$. The
two trees then report $g$ as $[0,1)$ and $\bot$ (no participation),
respectively. At the alternation $X$ in the first tree, the two candidates give
the end positions $(3,2,2)$ and $(3,3,1)$ for the root, $XY$, and $X$. They
first differ at $XY$, so the second candidate wins;
Section~\ref{sec:continuations} turns such lists into the comparison keys of
our algorithm.

Keep the tree $(XY)Z$ and vary only the length of the input. On \texttt{aa},
\texttt{aaa}, and \texttt{aaaa}, both branches of $X$ admit a full match each
time, yet the preferred $X$ is $aa$, then $a$, then $aa$ again
(Table~\ref{tab:context-priority}). What changes is how far $XY$ can reach.
Feasibility and a fixed priority between the branches of $X$ cannot explain
the three answers. Section~\ref{sec:related} uses this example to distinguish
our continuation comparisons from feasibility-guided greedy parsing.

\begin{table}[ht]
\centering\small
\begin{tabular}{@{}lccc@{}}
\toprule
Input & Latest end of $XY$ if $X=aa$ & Latest end of $XY$ if $X=a$ & Preferred $X$ \\
\midrule
\texttt{aa}   & $2$ & $1$ & $aa$ \\
\texttt{aaa}  & $2$ & $3$ & $a$ \\
\texttt{aaaa} & $4$ & $3$ & $aa$ \\
\bottomrule
\end{tabular}
\caption{The same two branches of $X$ are feasible on every input. Each end
position is maximized among full matches of $(XY)Z$ that use the specified
branch. POSIX first compares the parent $XY$, so neither fixed branch
priority gives all three preferred parses.}
\label{tab:context-priority}
\end{table}

An unparenthesized sequence such as $XYZ$ is read as $X(YZ)$, and explicit
parentheses are kept as written. With this reading, once the whole interval
and the earlier factors are fixed, the rest of the sequence has a fixed span,
so its first factor is compared next. This yields the left-to-right rule for
concatenation discussed in Austin Group issue~1857~\cite{austin1857}. For
instance, $XYZ$ selects the first decomposition of \texttt{aaa} listed above,
with $X=aa$.

\paragraph{An empty capture can participate.}
On the empty word, $(\capture{g}{a})?$ matches by skipping its body, so $g$
does not participate ($g=\bot$). With the operators swapped,
$\capture{g}{a?}$ matches with an empty body, so $g=[0,0)$. Neither reads a
letter. Section~\ref{sec:semantics} makes the distinction precise, including
for empty iterations of a repetition (Example~\ref{ex:empty-iteration}).

\section{Technical overview}
\label{sec:overview}

Three problems must be solved to select the POSIX parse in $O(m(n+1))$ time.
There are exponentially many parses; the keys that compare them can have
$\Theta(m)$ coordinates; and storing one choice per state costs $\Theta(mn)$
bits. We introduce the three solutions on one expression.

Consider $E:=(aa?)^*$. Write $E=F^*$ with body $F=a_1Q$, where $a_1$ is the
first literal and $Q=a_2?$ contains the second literal $a_2$. Although $a_1$
and $a_2$ read the same letter, they are distinct nodes of $E$. Each iteration
reads one or two letters, so $a^n$ has exponentially many full parses (the
Fibonacci number $f_{n+1}$). On \texttt{aaa} the three parses split the input
as $2+1$, $1+2$, and $1+1+1$, and POSIX selects $2+1$. How much of a parse must
an algorithm remember to make this decision?

\paragraph{Two histories, one continuation.}
We build a parse by a left-to-right depth-first traversal that enters nodes,
reads letters, and returns from nodes. Take the input \texttt{aaaa} and pause
after the second letter. The prefix \texttt{aa} may have been read by one
iteration or by two, so in both cases the traversal is about to close an
iteration of $F$ at position~$2$. In the first case that iteration started at
$0$, and in the second at $1$. In both cases, reading the remaining
\texttt{aa} as one iteration beats reading it as two, for the same reason: the
next iteration is longer. So both histories can share one decision about the
future. Lemma~\ref{lem:history-independence} shows that this holds in general:
the preferred continuation depends only on the current node, the current
position, and whether the node is being entered or left.

\paragraph{Comparing two continuations.}
Back to \texttt{aaa}, starting at $0$ and accepting any end position. After
$a_1$ reads a letter, $Q$ can stay empty or read the second \texttt{a}. Both
continuations can end the root at $3$. Their pairs
$(\text{end of }E,\ \text{end of the first iteration of }F)$ are $(3,1)$ and
$(3,2)$, so the second continuation wins at $F$. This comparison first examines
the outer node $E$, then its current iteration of $F$. In general, the
\emph{key} of a traversal state lists the end positions of all nodes that are
open at that state, from the root down. Thus the keys at $E$, $F$, $a_1$, $Q$,
and $a_2$ have $1$, $2$, $3$, $3$, and $4$ coordinates.

\paragraph{One layer per input position.}
For each node $e$ and position $i$ there are two states. In the entry state
$P_e(i)$, a nonempty (\emph{positive}) match of $e$ starts at $i$; in the
return state $C_e(i)$, such a match ends at $i$. We also write $P_e(i)$ and
$C_e(i)$ for the key of the preferred continuation from that state, or $\bot$
if there is none; $\bot$ is below every key. Keys are compared
lexicographically: at the first coordinate where they differ, the larger end
position wins. Let $\push_i$ append $i$, and let $\pop$ drop the last
coordinate; both leave $\bot$ unchanged.

The scan runs backward over the input, and layer $i$ computes all values
$C_e(i)$ and then all values $P_e(i-1)$. Table~\ref{tab:overview-trace} shows
the three layers on \texttt{aaa}. A state with two possible next actions is a
\emph{fork}. Here the forks are the returns from $F$ and from $a_1$, where
\[
\begin{aligned}
 C_F(i)&=\push_i\max\{C_E(i),\ \pop P_F(i)\},\\
 C_{a_1}(i)&=\push_i\max\{C_F(i),\ \pop P_Q(i)\}.
\end{aligned}
\]
At the return from $F$, stopping gives $C_E(i)$, while starting another
iteration gives the key of $P_F(i)$ without the coordinate of the new iteration;
hence the pop. At the return from $a_1$, the two candidates skip $Q$ or enter
it. Either way the node being left ends at $i$, so $\push_i$ appends $i$. The
other returns have a single candidate: $C_E(i)=(i)$, $C_Q(i)=\push_iC_F(i)$,
and $C_{a_2}(i)=\push_iC_Q(i)$. A literal that matches the letter before
position $i$ copies its return key at $i$ into its entry slot at $i-1$;
otherwise it writes $\bot$. Every other entry key at $i-1$ drops the last
coordinate of the entry key of its child.

\begin{table}[ht]
 \centering\small
 \begin{tabular}{@{}rllll@{}}
 \toprule
 $i$ & $C(i)$ & $P(i-1)$ & After $F$ & After $a_1$ \\
 \midrule
 3 & $3,33,333,333,3333$ & $3,33,333,333,3333$ & Stop & Empty $Q$ \\
 2 & $2,32,332,322,3222$ & $3,33,332,322,3222$ & Continue & Positive $Q$ \\
 1 & $1,31,321,311,3111$ & $3,32,321,311,3111$ & Continue & Positive $Q$ \\
 \bottomrule
 \end{tabular}
 \caption{Backward evaluation of $(aa?)^*$ on \texttt{aaa}, accepting every
 end position. The second and third columns list the keys at $E$, $F$, $a_1$,
 $Q$, and $a_2$, in this order; for instance, $321=(3,2,1)$. All $P(3)$ are
 $\bot$. The last two columns give the choices at the two forks.}
 \label{tab:overview-trace}
\end{table}

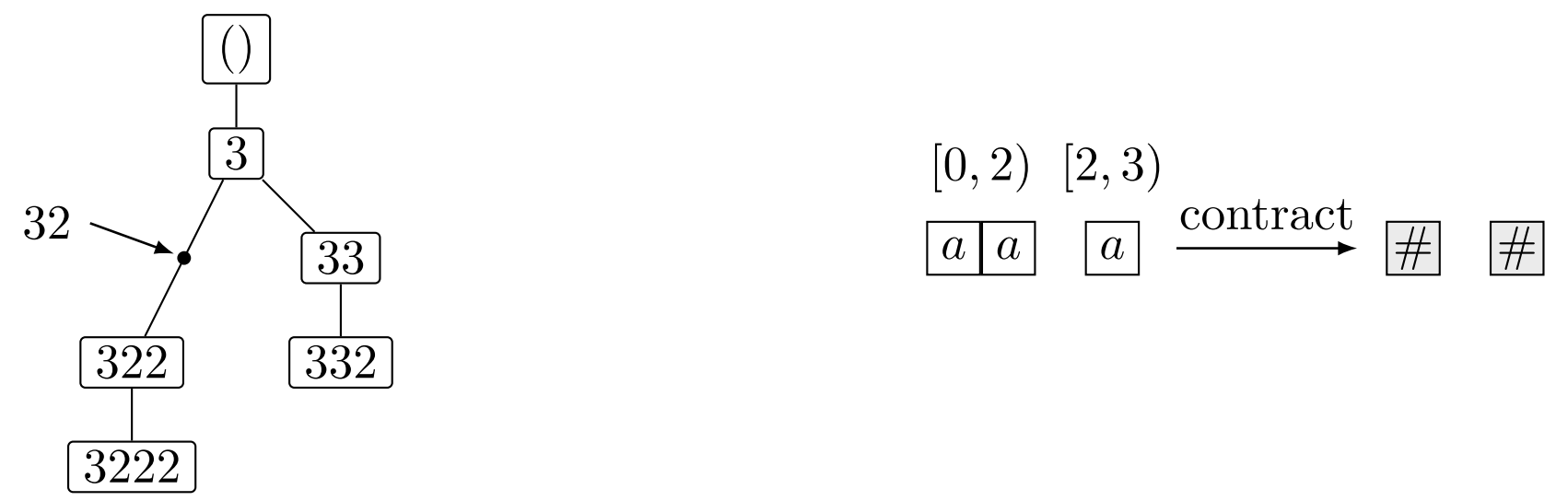
\begin{figure}[ht]
 \centering
 \begin{tikzpicture}[font=\small,x=1cm,y=1cm,
 key/.style={draw,rounded corners=1pt,inner xsep=3pt,inner ysep=2pt},
 token/.style={draw,minimum size=.38cm,inner sep=1pt},
 flow/.style={-{Latex[length=1.4mm]},line width=.5pt}]
 \node[key] (r) at (0,1.5) {$()$};
 \node[key] (k3) at (0,.75) {$3$};
 \node[key] (k322) at (-.75,-.75) {$322$};
 \node[key] (k3222) at (-.75,-1.5) {$3222$};
 \node[key] (k33) at (.75,0) {$33$};
 \node[key] (k332) at (.75,-.75) {$332$};
 \draw (r)--(k3) (k3)--(k322) (k322)--(k3222) (k3)--(k33) (k33)--(k332);
 \coordinate (mid) at ($(k3.south)!.5!(k322.north)$);
 \fill (mid) circle (1.4pt);
 \node[anchor=east] (lab) at (-1.05,.25) {$32$};
 \draw[flow] (lab.east)--($(mid)+(-.07,.03)$);
 \node[anchor=base] at (0,-2.35) {(a) The old keys at layer $1$};
 \node[token] (x0) at (5.15,.07) {$a$};
 \node[token,right=0pt of x0] (x1) {$a$};
 \node[token,right=10pt of x1] (x2) {$a$};
 \node[above=2pt of $(x0.north)!.5!(x1.north)$] {$[0,2)$};
 \node[above=2pt of x2] {$[2,3)$};
 \draw[flow] (6.75,.07)--node[above]{contract}(8.05,.07);
 \node[token,fill=black!8] (y0) at (8.45,.07) {$\#$};
 \node[token,fill=black!8,right=10pt of y0] {$\#$};
 \node[anchor=base] at (6.9,-2.35) {(b) Contract the selected $F$ spans};
 \end{tikzpicture}
 \caption{(a) The compressed trie of the old keys at the start of layer $1$ in
 Table~\ref{tab:overview-trace}. The prefix $32=\pop(322)$ lies inside the edge
 from $3$ to $322$. (b) The two selected iterations of $(aa?)^*$ on \texttt{aaa}
 give two tokens for the outer problem.}
 \label{fig:overview-operations}
\end{figure}

\paragraph{Comparing keys without copying them.}
A key can have $\Theta(m)$ coordinates, so copying or comparing keys explicitly
would cost $\Theta(m^2)$ time per layer. Consider how the last row of
Table~\ref{tab:overview-trace} is computed from the row above. To compute
$C_{a_1}(1)$, we compare $C_F(1)=31$ with $\pop P_Q(1)=\pop(322)=32$, and the
latter wins. The word $32$ is not a key of the previous layer but a prefix of
the old key $322$. We store the old keys in a trie and merge each path without
branches or stored keys into a single edge
(Figure~\ref{fig:overview-operations}a). The prefix $32$ then lies inside the
edge from $3$ to $322$. The trie stays intact throughout the layer: the lower
endpoint $322$ of the edge and the depth $2$ identify the prefix. Every
coordinate appended in layer $1$ is $1$, which is smaller than every old
coordinate. This observation lets us compare even such implicit prefixes in
constant time (Section~\ref{sec:evaluation}).

\paragraph{Recovering the iterations.}
Follow the choices in Table~\ref{tab:overview-trace} from $P_E(0)$. After the
first $a_1$, the choice of a positive $Q$ reads the second letter, and the
first iteration of $F$ ends at $2$. Another iteration starts, and after its
$a_1$ the option $Q$ stays empty and the repetition stops. Storing one bit per
state allows this forward replay but costs $\Theta(mn)$ bits. To use only
$O(m+\ell)$ words, we record just the visits to one chosen subtree, here $F$.
Its two iterations span $[0,2)$ and $[2,3)$. Replace each span by one copy of a
fresh symbol $\#$, and the subtree $F$ by a leaf that reads $\#$. This gives the
outer problem $\#^*$ on \texttt{\#\#} (Figure~\ref{fig:overview-operations}b).
Its two marker leaves refer to the inner solutions for $F$ on \texttt{aa} and on
\texttt{a}. Section~\ref{sec:outputs} explains why this decomposition preserves
the selected tree.

\section{Specification}\label{sec:semantics}

A query selects an interval, a parse tree on that interval, and the captures
read from that tree. We first specify the candidate trees and their order.

\paragraph{Expressions and positions.}
\label{sem:def-native}\label{sem:def-language}
Fix a finite alphabet $\Sigma$. The root expression $E$ and its subexpressions
have the following grammar.\footnote{The bar inside $(e_0\mid e_1)$ denotes
alternation. The other bars separate grammar productions.}
\[
 e ::= \emptyre\mid\eps\mid a\mid e_0e_1\mid(e_0\mid e_1)
       \mid e^*\mid e^+\mid e?\mid\capture{g}{e},
 \qquad a\in\Sigma.
\]
We call the $m$ nodes of the syntax tree of $E$ simply its \emph{nodes}; one
subexpression may occur at several nodes, as the two literals of $(aa?)^*$ do.
Each capture has a distinct name $g$, and the name $0$ is reserved for the whole
match. Write $L(e)$ for the language of a node $e$; a capture has the language
of its child, and $e$ is \emph{nullable} if $\eps\in L(e)$. We write
$\RepExpr_{\lo,\hi}(e)$ for any of the three quantifiers, with lower and upper
counts $(\lo,\hi)$ equal to $(0,\infty)$, $(1,\infty)$, or $(0,1)$ for star,
plus, or option, respectively.

The text $w$ has length $n$. Its \emph{positions} are $0,\ldots,n$, and for
$i<n$, $w[i]$ is the character between positions $i$ and $i+1$. For
$0\le i\le j\le n$, the interval $[i,j)$ holds the characters
$w[i],\ldots,w[j-1]$, which form the substring $w[i:j)$.

\begin{definition}[Canonical parses]\label{sem:def-interval-parses}
\label{sem:def-canonical}\label{sem:def-candidates}\label{cap:spans}
A \emph{canonical parse} of a node $e$ on $[i,j)$ is an ordered tree whose
vertices each record a node of $E$ and an interval. Its root records $e$ and
$[i,j)$, and its children are canonical parses of children of $e$, as follows.
\begin{enumerate}[label=(\roman*),leftmargin=*]
\item A literal $a$ reads one $a$, and $\eps$ reads the empty word. Both are
      leaves. There is no parse of $\emptyre$.
\item A concatenation has two children, on $[i,k)$ and $[k,j)$. An alternation
      records its branch and has one child on $[i,j)$, a parse of that branch.
      A capture has one child on $[i,j)$.
\item A repetition $\RepExpr_{\lo,\hi}(e)$ has between $\lo$ and $\hi$
      consecutive children, which are parses of its body $e$ and together span
      $[i,j)$. If there are at least two, each has positive length. Zero
      children require $i=j$.
\end{enumerate}
\end{definition}

We call the vertices of a parse $T$ \emph{visits}: a node of $E$ can be visited
several times, for instance once per iteration of a repetition. The
\emph{body} of a repetition is its child node, and an \emph{iteration} is a
visit to the body. A visit or a match is \emph{positive} if its interval is
nonempty. If $T$ spans $[i,j)$, write $\spanop(T):=[i,j)$ and $\len(T):=j-i$.
Denote by $\Parse_e^w(i,j)$ the set of canonical parses of $e$ on $[i,j)$.
Canonicality restricts the candidates before any preference is applied.

\begin{example}[Empty iterations]\label{ex:empty-iteration}
On the empty word, $(\capture{g}{a?})^*$ has exactly two canonical parses:
zero iterations, which leave $g$ unset, and one empty iteration, which visits
$g$ on $[0,0)$. The list of two empty iterations is not canonical.
Definition~\ref{sem:def-recursive-order} below prefers the single iteration, so
$g=[0,0)$. This follows Okui and Suzuki~\cite[Definition~3]{okui2013}; Urban's
positive-iteration convention~\cite{urban2023} allows only the empty list here,
so $g$ is unset in his semantics.
\end{example}

\begin{definition}[POSIX preference]\label{sem:def-recursive-order}
\label{sem:def-positional-order}
For parses of the same node $e$ with the same start, write $T\succ_e U$ if $T$
is preferred to $U$ by the following tests, applied in order.
\begin{enumerate}[label=(\roman*),leftmargin=*]
\item Prefer the longer root interval.
\item At an alternation with different branches, prefer the left branch.
\item Compare the ordered lists of children lexicographically by this same
      preference. A proper extension wins over its prefix.
\end{enumerate}
Write $T\succeq_e U$ for $T\succ_e U$ or $T=U$.
\end{definition}

Equal preceding children read the same text, so each recursive comparison
starts at a common position. For example, the three parses of
$(\capture{g}{a\mid aa})^*$ on \texttt{aaa} in Table~\ref{tab:three-parses} all
span $[0,3)$. In table order, their first iterations end at $1$, $1$, and $2$,
so $aa/a$ is preferred. The first two rows agree on the first iteration, and
their second iterations end at $2$ and $3$, respectively. Hence
$aa/a\succ a/aa\succ a/a/a$.

To compare continuations in Section~\ref{sec:continuations}, we need a linear
form of this recursive order. The \emph{code} of a parse $T$ that ends at $j$
and has the children $T_1,\ldots,T_k$ is
\[
 \codeop(T):=\opentok(j)\ \tagop(T)\ \codeop(T_1)\cdots\codeop(T_k)\ \closetok,
\]
where the tag $\tagop(T)$ is $\mathsf L$ or $\mathsf R$ for the branch of an
alternation and empty otherwise. Tokens are ordered by
\[
 \closetok<\opentok(j)<\opentok(j')\ \text{ for }j<j',\qquad
 \mathsf R<\mathsf L,
\]
and codes are compared lexicographically; at corresponding positions a tag
always meets a tag. For $a^*a^*$ on \texttt{a}, the parse that gives the letter
to the first star and the parse that gives it to the second star have the codes
\[
\begin{aligned}
 &\opentok(1)\ \opentok(1)\ \opentok(1)\ \closetok\ \closetok\ \opentok(1)\
  \closetok\ \closetok,\\
 &\opentok(1)\ \opentok(0)\ \closetok\ \opentok(1)\ \opentok(1)\ \closetok\
  \closetok\ \closetok.
\end{aligned}
\]
The second token decides, just as Definition~\ref{sem:def-recursive-order}
decides at the left child.

\begin{lemma}[Depth-first code]\label{lem:parse-code}
For parses $T$ and $U$ of a common node $e$ with a common start,
$T\succ_e U$ if and only if $\codeop(T)>_{\mathrm{lex}}\codeop(U)$.
\end{lemma}
\begin{proof}
At a pair of corresponding visits the starts agree, so their end positions
compare in the same order as their lengths. If the ends agree, the branch tags,
when present, precede the children. A difference within a child is met before
any token of a later child. If the common children run out on one side only,
that side emits a closing token before the other side's next opening token.
Thus only the comparisons of Definition~\ref{sem:def-recursive-order} can
decide. Equal codes determine the same tree, given $e$ and the common start.
\end{proof}

\begin{lemma}[Feasibility and preference]\label{sem:finite}
\label{sem:order-total}
The set $\Parse_e^w(i,j)$ is finite, and it is nonempty exactly when
$w[i:j)\in L(e)$. For a fixed node and start, the preference is a strict total
order on the candidates.
\end{lemma}
\begin{proof}
Induct on $e$. For a repetition with a positive iteration, remove all empty
iterations from a witness of membership. The remaining count is at least one
and at most $\hi$, so it is allowed since $\lo\le1$. If all iterations are
empty, the repetition can use zero iterations when $\lo=0$, and a single empty
iteration otherwise. This proves the equality of languages, using the
induction hypothesis for each child. Finiteness follows because there are
finitely many cuts and a positive repetition on $[i,j)$ has at most $j-i$
iterations. Totality follows from Lemma~\ref{lem:parse-code}.
\end{proof}

For a feasible interval, $\Opt_e^w(i,j)$ denotes the unique candidate $T$ with
$T\succeq_e U$ for every $U\in\Parse_e^w(i,j)$, the \emph{preferred} parse. Our
algorithm inserts each empty subtree in a single step, and the next two facts
justify this.

\begin{lemma}[Same-interval replacement]\label{sem:replacement}
Replacing a subtree of a canonical parse by a parse of the same node on the same
interval that is preferred to it gives a canonical parse that is preferred to
the original.
\end{lemma}
\begin{proof}
The replacement preserves the numbers of children and all interval constraints.
At each proper ancestor, the comparison reaches the changed subtree before any
later child, so the improvement reaches the root.
\end{proof}

For a nullable node $e$, let $\nu_i(e)$ be its preferred parse on the empty
interval $[i,i)$, the \emph{preferred empty parse}. Its shape, which we call
the \emph{template} of $e$, does not depend on $i$. We store each template as
a constant-size record at its node, with references to the templates of its
selected children. Thus the templates share their subtrees.

\begin{lemma}[Preferred empty parses]\label{lem:empty-template}
All nullability bits and templates can be computed in $O(m)$ time and words.
Every empty subtree of a preferred parse is the corresponding $\nu_i(e)$.
\end{lemma}
\begin{proof}
Compute nullability bottom-up. An empty parse takes both children of a
concatenation, the first nullable branch of an alternation, and the child of a
capture. A nullable repetition takes one empty iteration if its body is
nullable, and no iteration otherwise; one iteration wins the comparison of
lists. These local rules determine each template by induction on $e$ and need
at most two child references per node, so all templates take $O(m)$ time and
words. Expanding one template visits each included node once and gives it the
interval $[i,i)$. The claim about preferred parses follows from
Lemma~\ref{sem:replacement}.
\end{proof}

\begin{definition}[Queries]\label{sem:def-query}
\label{sem:def-optimum}\label{sem:def-prefix}\label{alg:search-semantics}
Let $J\subseteq\{0,\ldots,n\}$ be a set of permitted end positions with
constant-time membership tests. The \emph{answer} at a start $i$ is
\[
 A_J(i):=\bigl(j^\star,\ \Opt_E^w(i,j^\star)\bigr),\qquad
 j^\star:=\max\{\,j\in J:\ j\ge i\text{ and }\Parse_E^w(i,j)\ne\emptyset\,\},
\]
or \emph{failure} if no such $j$ exists. Full matching asks for
$A_{\{n\}}(0)$. Longest-prefix matching asks for $A_J(i)$ at a given start $i$,
and leftmost-longest search for $A_J(i)$ at the least start $i$ whose answer is
not a failure, both with $J=\{0,\ldots,n\}$.
\end{definition}

\begin{definition}[Capture vector]\label{cap:forest}\label{cap:projection}
Add to a parse $T$ a synthetic capture $0$ around the whole match. For a capture
$g$, let $\parcap(g)$ be the nearest proper ancestor of $g$ in $E$ that is a
capture, or $0$ if there is none. Likewise, the \emph{capture parent} of a
visit to $g$ is its nearest proper ancestor visit to a capture, or the synthetic
root. For a visit $v$, let $\lastv_T(g,v)$ be the last visit to $g$ in
depth-first order whose capture parent is $v$, or $\bot$ if there is none, and
set $\lastv_T(g,\bot):=\bot$. The \emph{reported visits} are
\[
 \rep_T(0):=\text{the synthetic root},\qquad
 \rep_T(g):=\lastv_T\bigl(g,\rep_T(\parcap(g))\bigr).
\]
The \emph{capture vector} of $T$ consists of the spans
$\capv_T(g):=\spanop(\rep_T(g))$ of all captures $g$, including $0$, where
$\spanop(\bot):=\bot$.
\end{definition}

This is the reporting rule of \texttt{regexec}~\cite{posix2024regexec}: a group
that matches several times reports its last match, and a group nested in another
is reported only within the reported match of the outer group. An empty group
that participates reports $[i,i)$, while a group absent from the reported match
of its parent reports $\bot$.

\begin{example}[Nested captures]\label{ex:nested-captures}
Let $E=(\capture{k}{\capture{g}{a}\mid b})^*$ and $w=\texttt{ab}$. The preferred
parse has two iterations. $\rep_T(k)$ is the second visit to $k$, on $[1,2)$,
and no visit to $g$ has it as capture parent. Hence $\capv_T(g)=\bot$, although
$g$ matched $[0,1)$ in the first iteration. The same happens to an empty $g$ at
the boundary between iterations. In $(\capture{k}{a\capture{g}{\eps}\mid b})^*$
on \texttt{ab}, the only visit to $g$ is on $[1,1)$ inside the first iteration,
so $g$ is unset, although the reported iteration also starts at $1$.
\end{example}

For a capture $g$, let $\lastv_T(g)$ be its last visit in depth-first order, or
$\bot$ if there is none, and let $\lastv_T(0)$ be the synthetic root. The
reported visit is then easy to find.

\begin{lemma}[Last visit under the reported parent]\label{out:last-visit}
Let $p=\parcap(g)$. Then $\rep_T(g)=\lastv_T(g)$ if $\rep_T(p)\ne\bot$ and
$\lastv_T(g)$ lies in the subtree of $\lastv_T(p)$; otherwise
$\rep_T(g)=\bot$.
\end{lemma}
\begin{proof}
Induct in parent-before-child order. By induction, a reported parent is its own
last visit. Every visit to $g$ lies in some visit to $p$, and these subtrees are
disjoint and ordered in depth-first order. If the last visit to $p$ contains a
visit to $g$, it contains the last visit to $g$ overall. Otherwise
Definition~\ref{cap:projection} gives $\bot$.
\end{proof}

\subsection{Correspondence with positional POSIX orders}
\label{sem:correspondence}

The comparison of Definition~\ref{sem:def-recursive-order} can also be expressed
by comparing lengths at ordered tree addresses, as in two established
conventions. Both conventions ignore captures; since the syntax tree is fixed,
the captures can be restored with their intervals and nesting.

\begin{proposition}[Scope of the correspondence]
\label{sem:correspondence-scope}
For a fixed node and interval, the following hold.
\begin{enumerate}[label=(\roman*),leftmargin=*]
\item If star is the only quantifier, canonical parses correspond bijectively
and in preference order to the canonical trees of Okui and
Suzuki~\cite{okui2013}.
\item With native $*$, $+$, and $?$, the POSIX preference agrees on canonical
parses with the positional P-order of Borsotti and
Trofimovich~\cite{borsotti2021}. Admitting their additional empty iterations
gives the same optimum whenever a parse exists.
\end{enumerate}
\end{proposition}
\begin{proof}
Proposition~\ref{sem:external-agreement} in Appendix~\ref{sec:equivalence} gives the
order-preserving translations.
\end{proof}

\section{Comparing continuations}
\label{sec:continuations}\label{sec:algorithm}

To avoid keeping every history that reaches a state, we determine which parts
of a history can affect later choices. Fix the text $w$ and the set $J$ of
permitted end positions. Call a parse \emph{empty-normalized} if every empty
visit to a node $e$ at a position $i$ is the preferred empty parse $\nu_i(e)$.
By Lemma~\ref{lem:empty-template} the preferred parse is empty-normalized. We
can therefore traverse the positive visits one by one and insert each empty
subtree in a single step.

\subsection{The traversal graph}

For each node $e$ and position $i$, the \emph{entry state} $P_e(i)$ begins a
positive match of $e$ at $i$, and the \emph{return state} $C_e(i)$ ends one at
$i$. At either state, the open visits are, from the outermost, visits to the
nodes on the path from the root to $e$. The \emph{traversal graph} $G_w$ has
these states, an accepting state $\acc$, and the transitions of
Table~\ref{tab:continuation-actions}. A transition labelled $\nu_i(x)$ inserts
the empty parse $\nu_i(x)$ as a child of the current visit.

\begin{table}[H]
\centering\small
\begin{tabular}{@{}lll@{}}
\toprule
Transition & Condition & Action\\
\midrule
$P_e(i)\to P_f(i)$ & $f$ is the first or only child of $e$ & enter $f$\\
$P_{fg}(i)\xrightarrow{\nu_i(f)}P_g(i)$ & $f$ is nullable & enter $g$ after an empty $f$\\
$P_{f\mid g}(i)\to P_g(i)$ & & enter the second branch\\
$P_a(i)\to C_a(i+1)$ & $i<n$ and $w[i]=a$ & read a letter\\
\midrule
$C_f(i)\xrightarrow{\nu_i(g)}C_{fg}(i)$ & $g$ is nullable & return with an empty $g$\\
$C_f(i)\to P_g(i)$ & $f$ is the left child of $fg$ & enter $g$\\
$C_f(i)\to C_p(i)$ & $f$ is a child of $p$, but not the left child of a concatenation & return to $p$\\
$C_f(i)\to P_f(i)$ & $f$ is the body of $f^*$ or $f^+$ & start another iteration\\
$C_E(i)\to\acc$ & $i\in J$ & accept\\
\bottomrule
\end{tabular}
\caption{The transitions of the traversal graph $G_w$. The transitions from an
entry state are listed in tie-breaking order
(Lemma~\ref{lem:fork}).}
\label{tab:continuation-actions}
\end{table}

\begin{lemma}[Traversals and parses]\label{lem:traversals}
The graph $G_w$ is acyclic. For every start $i$, the paths of $G_w$ from $P_E(i)$
to $\acc$ correspond one-to-one to the positive empty-normalized parses of $E$
that start at $i$ and end in $J$.
\end{lemma}
\begin{proof}
At a fixed position, returns move up the syntax tree and may switch to entries,
which only move down. Only reading a letter leads from an entry back to a
return, and it advances the position. Hence $G_w$ is acyclic. A positive
concatenation starts with its left child if that child is positive, and
otherwise with its right child. A positive option has one positive iteration,
and a positive star or plus has one or more. Alternations and captures pass
positivity to the chosen child. Induction on $E$ thus gives a unique path for
each such parse, including its empty subtrees, and every path arises in this
way.
\end{proof}

For $(aa?)^*$ on \texttt{aaa} (Section~\ref{sec:overview}), the preferred parse
is the path
\[
\begin{aligned}
&P_E(0)\to P_F(0)\to P_{a_1}(0)\to C_{a_1}(1)\to P_Q(1)\to P_{a_2}(1)\to
C_{a_2}(2)\to C_Q(2)\to C_F(2)\\
&\quad\to P_F(2)\to P_{a_1}(2)\to C_{a_1}(3)\xrightarrow{\nu_3(Q)}C_F(3)\to
C_E(3)\to\acc.
\end{aligned}
\]

Like the greedy parser of Frisch and
Cardelli~\cite[Section~5]{frischcardelli2004}, our states are locations in the
syntax tree. Their parser adds a progress flag for repetitions of nullable
bodies; our entry and return states admit only positive visits, and empty
subtrees are inserted as templates.

\subsection{The order of continuations}

A \emph{continuation} $\rho$ from a state $\sigma$ is a path from $\sigma$ to
$\acc$. At either $\sigma=P_e(i)$ or $\sigma=C_e(i)$, regard the root-to-$e$
path as a stack of open visits, even if no path from a root entry reaches
$\sigma$. The actions of $\rho$ determine the end of each visit in this stack.
The \emph{key} $K(\rho)$ lists these end positions from the root down and has
$D(e)$ coordinates, where $D(e)$ is the number of nodes on that path.

Interpret the actions as emitting the code of Lemma~\ref{lem:parse-code}.
The opening tokens of the initially open visits have already been emitted,
with end positions to be filled by $K(\rho)$. Let $W(\rho)$ be the remaining
tokens: entering a visit emits its opening token, choosing a branch emits its
tag before entering the child, leaving a visit emits its closing token, and
inserting an empty subtree emits its whole code. Write $\rho\succ_\sigma\rho'$
if $K(\rho)>_{\mathrm{lex}}K(\rho')$, or if the keys agree and
$W(\rho)>_{\mathrm{lex}}W(\rho')$. This defines continuation preference at
every entry and return state. At $\acc$, the unique continuation is the empty
path, with empty key and suffix; thus $\succ_{\acc}$ is the empty strict order.

A \emph{history} $h$ is a path from a root entry $P_E(i_0)$ to $\sigma$.
Write $T_h(\rho)$ for the parse formed by $h$ followed by $\rho$.
In Section~\ref{sec:overview}, the two histories of \texttt{aaaa} that reach
$C_F(2)$, one through the iteration $[0,2)$ and one through $[0,1)$ and $[1,2)$,
form such a pair $h,h'$.

\begin{lemma}[History independence]\label{lem:history-independence}
For every history $h$ ending at $\sigma$ and continuations $\rho,\rho'$ from
$\sigma$,
\[
 \rho\succ_\sigma\rho'\quad\Longleftrightarrow\quad
 T_h(\rho)\succ_E T_h(\rho').
\]
Moreover, let $a$ be a legal action from $\sigma$ to $\sigma'$, and write
$a\rho$ for the continuation that takes $a$ and then follows $\rho$. Then, for
all continuations $\rho,\rho'$ from $\sigma'$,
\[
 a\rho\succ_\sigma a\rho'\quad\Longleftrightarrow\quad
 \rho\succ_{\sigma'}\rho',
\]
whether or not $\sigma$ is reachable from a root entry.
\end{lemma}
\begin{proof}
The legal next actions depend only on the state. In particular, a positive
iteration of a star or plus can stop or repeat after any number of earlier
iterations. Thus every continuation from $\sigma$ is legal after $h$.

Consider the codes of Lemma~\ref{lem:parse-code}. Among the tokens emitted along
$h$, only the end positions in the opening tokens of the visits open at $\sigma$
are undetermined. The continuation $\rho$ fills these fields with $K(\rho)$, in
order, and appends $W(\rho)$. All other tokens of the history agree
in $T_h(\rho)$ and $T_h(\rho')$, and the opening tokens of the open visits occur
from the root down. Hence the comparison of the two codes first compares
$K(\rho)$ with $K(\rho')$ and, on equality, $W(\rho)$ with $W(\rho')$. Neither
comparison depends on $h$. Since each parse has a unique path
(Lemma~\ref{lem:traversals}), this also gives an order isomorphism between the
parses that complete any two histories ending at $\sigma$.

For the second claim, read the fields of a continuation as the coordinates of
its key followed by the tokens of its suffix. Prepending $a$ inserts fields
that are the same for $\rho$ and $\rho'$: when $a$ leaves a node at $i$, the
end position $i$ and a closing token, and in addition a branch tag or the code
of an inserted empty subtree. The only other change concerns a visit that $a$
enters. Its end position moves from the last coordinate of the key of $\rho$ to
its opening token, which still follows the end positions of all shared
ancestors and precedes the rest of the suffix. Hence $a\rho$ and $a\rho'$
first differ in the field where $\rho$ and $\rho'$ first differ, in the same
direction. This argument uses only the action and the root-to-node stacks, not
a history.
\end{proof}

A \emph{fork} is a state with two possible next actions. When two continuations
from a fork have equal keys, their first actions decide.

\begin{lemma}[Comparison at a fork]\label{lem:fork}
Let $\rho$ and $\rho'$ be continuations from a fork $\sigma$ whose first actions
differ. If $K(\rho)\ne K(\rho')$, then $\rho\succ_\sigma\rho'$ if and only if
$K(\rho)>_{\mathrm{lex}}K(\rho')$. If $K(\rho)=K(\rho')$, then the fork is an
entry state, and $\rho\succ_\sigma\rho'$ if and only if the first action of
$\rho$ precedes that of $\rho'$ in Table~\ref{tab:continuation-actions}.
Consequently, continuation preference is a strict total order on the
continuations from any entry or return state.
\end{lemma}
\begin{proof}
The first claim is the definition of continuation preference. Suppose that the
keys agree. At an alternation, the next differing token is the branch tag, and the left branch
wins. At the entry of a concatenation, the next differing token is the end of
its left child. That child starts at the current position in both candidates
and ends strictly later in the candidate that enters it as a positive visit.
At the return from the left child of a concatenation or from an iteration, the
parent ends at the current position if we return at once, and strictly later if
a positive visit follows. The keys then already differ at the coordinate of the
parent, so equal keys occur only at entry states.

Continuation preference compares $(K,W)$ lexicographically, so it is
irreflexive and transitive. Two distinct continuations from a state first
differ in their actions at some fork. The claims above order them there, and
the second claim of Lemma~\ref{lem:history-independence} carries this order
back through their common actions. Hence any two distinct continuations are
comparable.
\end{proof}

For example, let $E:=\capture{g}{a}\mid a$ and $w:=\texttt{a}$, with
$J=\{1\}$. Both continuations from $P_E(0)$ have key $(1)$, but the left
branch wins the tie and reports $g=[0,1)$; the right branch would report
$g=\bot$. Thus equal keys can still lead to different capture outputs.

\subsection{Two update rules}

We also write $P_e(i)$ and $C_e(i)$ for the \emph{values} of these states: the
key of the preferred continuation, or $\bot$ if there is none; $\bot$ is below
every key. As in Section~\ref{sec:overview}, $\push_i$ appends the coordinate
$i$ and $\pop$ drops the last coordinate; both map $\bot$ to $\bot$. For an
ordered list of candidates, $\firstmax$ returns the first one
with the maximum key, or $\bot$ if all are $\bot$, and the scan keeps the chosen
action beside the value. For a node $e$ that is not a literal, let
$\mathcal I(e)$ list the children of $e$ that an entry can enter, in the order
of Table~\ref{tab:continuation-actions}. For a node $e$ with parent $p$, let
$\mathcal R_e(i)$ list $C_p(i)$ if a return to $p$ is legal, and $\pop P_x(i)$ if
entering a positive visit to $x$ is legal, where $x$ is the right sibling of $e$
or $e$ itself. The scan computes
\begin{align}
 P_e(i) &= \firstmax\bigl(\pop P_f(i): f\in\mathcal I(e)\bigr),
 \label{eq:entry-update}\\
 C_e(i) &= \push_i\bigl(\firstmax(\mathcal R_e(i))\bigr).
 \label{eq:return-update}
\end{align}
At a literal $e$ with the letter $a$, $P_e(i)=C_e(i+1)$ if $i<n$ and $w[i]=a$, and
$P_e(i)=\bot$ otherwise. At the root, $C_E(i)=(i)$ if $i\in J$, and
$C_E(i)=\bot$ otherwise.

The pop in \eqref{eq:entry-update} removes the coordinate of the child being
entered, so the candidates at an entry may have equal keys. For example, for
$a?a^*$ on \texttt{a} with $J=\{1\}$, both candidates at the root entry have the
key $(1)$, and the positive left child wins by Lemma~\ref{lem:fork}.

\begin{proposition}[Correctness of the update rules]\label{alg:correct}
For every entry or return state, the update rules give the key and the first
action of its preferred continuation, or $\bot$ if
there is none. Consequently, if $P_E(s)\ne\bot$, following the chosen actions
from $P_E(s)$ yields $A_J(s)$. Otherwise $A_J(s)=(s,\nu_s(E))$ if $E$ is nullable
and $s\in J$, and $A_J(s)$ is a failure if not.
\end{proposition}
\begin{proof}
Proceed in reverse topological order of $G_w$. Fix a legal next action from a
state. By Lemma~\ref{lem:history-independence}, the preferred continuation of
its successor remains preferred when prefixed by this action. On
an entry, the successor opens a child, and dropping the coordinate of that child
gives a key for the current state. On a return, both candidates describe the
open visits up to the parent, and the current node ends at $i$, so $\push_i$
appends $i$. Reading a letter leaves the open visits unchanged.
Lemma~\ref{lem:fork} then selects the preferred successor, with acceptance at
the root as the base case. At a fixed start a positive root parse is preferred
to the empty one, and Lemma~\ref{lem:empty-template} gives the empty parse in
the remaining case.
\end{proof}

\begin{lemma}[Ranges of the coordinates]\label{lem:key-range}
Every value $C_e(i)\ne\bot$ is nonincreasing and ends with $i$. Every value
$P_e(i)\ne\bot$ is nonincreasing, and all its coordinates exceed $i$.
\end{lemma}
\begin{proof}
Induct in reverse topological order; the claims also hold at states that no
history reaches. The literal rule copies a return value at $i+1$, and an entry
only drops coordinates. A return appends $i$ to a candidate whose coordinates
are at least $i$, since $C_p(i)$ ends with $i$ and $\pop P_x(i)$ has coordinates
above $i$.
\end{proof}

\begin{remark}[Shorter keys]\label{rem:reduced-keys}
A child of an alternation or a capture, the body of an option, and the right
child of a concatenation always end together with their parent. Their
coordinates can therefore be omitted from the keys: repeating each remaining
coordinate the number of times prescribed by a fixed state's root-to-node path
recovers the full key. The omission therefore preserves comparisons between
candidates at that state. It can shorten keys and save some pops; on $(aa?)^*$
it leaves two coordinates at $Q$ and at $a_2$. The bounds do not change, and we
use full keys throughout.
\end{remark}

\section{Linear-time evaluation of each layer}
\label{sec:evaluation}\label{sec:compact-keys}

A key can have as many coordinates as the depth of $E$, which can be
$\Theta(m)$. To spend $O(m)$ time per position, we need $O(m)$ key operations
per layer (Section~\ref{sec:schedule}), each in constant time, plus one $O(m)$
rebuild per layer (Sections~\ref{sec:key-order} and~\ref{sec:frozen}). Constant
time relies on the fact that all coordinates created in one layer are equal.

\subsection{Two passes per layer}\label{sec:schedule}

Layer $i$, for $n\ge i\ge1$, turns the array $A[e]=P_e(i)$ into
$A[e]=P_e(i-1)$. It first computes $C_e(i)$ for all nodes in preorder, and then
$P_e(i-1)$ for all nodes in postorder (Figure~\ref{fig:fused-visit}).

\begin{figure}[H]
\centering
\begin{minipage}{.9\linewidth}
\small
\begin{tabbing}
\hspace{1em}\=\hspace{1em}\=\hspace{1em}\=\kill
\textbf{initialize} $A[e]\gets\bot$ for every node $e$ \quad (these are the values $P_e(n)$)\\
\textbf{inspect} $A[E]$ at start $n$\\
\textbf{for} $i=n,n-1,\ldots,1$\\
\> \textbf{for} every node $e$ in preorder: compute $C_e(i)$ by the root rule
   or \eqref{eq:return-update}\\
\> \textbf{for} every node $e$ in postorder: $A[e]\gets P_e(i-1)$ by
   \eqref{eq:entry-update}, or by the literal rule from $C_e(i)$\\
\> \textbf{inspect} $A[E]$ at start $i-1$\\
\> rebuild the trie of the keys in $A$ for layer $i-1$
   (Lemma~\ref{lem:frozen-rebuild})
\end{tabbing}
\end{minipage}
\caption{The scan. Inspecting $A[E]$ at a start $s$ reads the end position of
the answer if $A[E]\ne\bot$. Otherwise the answer is $[s,s)$ if $E$ is nullable
and $s\in J$, and failure if not.}
\label{fig:fused-visit}
\end{figure}

The scan evaluates $P_e(i)$ for $0\le i\le n$ and $C_e(i)$ for
$1\le i\le n$. Return states at position $0$ are unnecessary: a path from a
root entry first reaches a return state by reading a letter, and positions
never decrease. Empty answers are handled directly by $\nu_s(E)$.

\begin{lemma}[Evaluation order]\label{alg:schedule}\label{alg:in-place}
The scan of Figure~\ref{fig:fused-visit} evaluates the updates of
Section~\ref{sec:continuations} in dependency order, with $O(m)$ key operations
per layer.
\end{lemma}
\begin{proof}
No positive visit starts at $n$, so the initialization gives all values
$P_e(n)$. In layer $i$, the root rule first gives $C_E(i)$. For every other
node $e$, the preorder pass computes $C_e(i)$ from $C_p(i)$, which it has already
computed, and from old values $P_x(i)$, which the second pass has not yet
overwritten. The postorder pass computes $P_e(i-1)$ from the new values
of the children of $e$, or from $C_e(i)$ at a literal. Each node takes a
constant number of key operations in each pass.
\end{proof}

\subsection{The order of keys within a layer}\label{sec:key-order}

By Lemma~\ref{lem:key-range}, all coordinates in $A$ exceed $i$ at the start of
layer $i$, and the layer only appends the coordinate $i$. From now on we read a
key as a word whose symbols are its coordinates. Every key created in layer $i$
thus has the form $\alpha i^k$: an \emph{old prefix} $\alpha$, possibly empty,
followed by $k\ge0$ copies of the \emph{fresh symbol} $i$. The order of two old
prefixes suffices to order their extensions.

\begin{lemma}[Fresh minimum]\label{lem:fresh-minimum}
Order words lexicographically, with a proper prefix first. If $x$ is smaller
than every symbol of $\alpha$ and $\beta$, and $k,k'\ge0$, then
\[
 \alpha x^k<_{\rm lex}\beta x^{k'}
 \quad\Longleftrightarrow\quad
 \alpha<_{\rm lex}\beta\ \text{ or }\ (\alpha=\beta\ \text{ and }k<k').
\]
\end{lemma}
\begin{proof}
A difference before either old word ends is unaffected by the appended
symbols. If one old word is a proper prefix of the other, the shorter word
either ends there or continues with $x$, which is smaller than the next symbol
of the longer old word; this also covers empty words. Equal old words reduce
the comparison to the numbers of appended symbols.
\end{proof}

Store the \emph{old keys}, the values in $A$ at the start of the layer, in a
\emph{trie}: the tree of their prefixes, in which an edge labelled $y$ extends a
word $\alpha$ to $\alpha y$, and the children of a vertex are ordered by
increasing symbol. Keep the root, every vertex at which an old key ends, and
every vertex with at least two children, and suppress the others. In this
\emph{compressed trie}, the word $\lambda(v)$ on the path from the root to a
vertex $v$ has length $\tdepth(v)$. An edge is stored by its endpoints and its
position among its siblings; its word is never stored. Each leaf ends an old
key, and there are at most $m$ old keys, so the trie has $O(m)$ vertices.

The trie stays fixed throughout the layer. Let $\pre(v)$ be the preorder number
of a vertex $v$. For an old prefix $\alpha$, let $v(\alpha)$ be its vertex if it
has one, and otherwise the lower endpoint of the compressed edge that contains
it; the empty prefix uses the root.

\begin{lemma}[Prefix order]\label{lem:prefix-order}
For old prefixes $\alpha$ and $\beta$, $\alpha<_{\rm lex}\beta$ if and only if
$\bigl(\pre(v(\alpha)),|\alpha|\bigr)<\bigl(\pre(v(\beta)),|\beta|\bigr)$
lexicographically.
\end{lemma}
\begin{proof}
The preorder of the uncompressed trie visits the prefixes in lexicographic
order. The prefixes inside a compressed edge, together with its lower endpoint,
form one consecutive block of that preorder, in increasing length, and the
preorder of the lower endpoints orders the blocks.
\end{proof}

\subsection{A frozen trie and two fields per key}\label{sec:frozen}

A \emph{handle} represents a key $\alpha i^{D-b}$ of length $D$ by two integers:
the identifier of $v:=v(\alpha)$ and $b:=|\alpha|$. Its word is
\[
          \lambda(v)[0:b)\, i^{D-b}.
\]
The length $D$ is not stored: a value at $P_e(i)$ or $C_e(i)$ has $D(e)$
coordinates, and a popped value has the length at the parent. A push therefore
keeps the handle, and so does a pop that removes a fresh symbol. Each slot stores
its handle together with the chosen action and the end position of the root in
the preferred continuation, which the scan copies from the chosen successor.

\begin{lemma}[Handle operations]\label{lem:handles}
On handles, comparing two keys of equal length, appending the fresh symbol, and
dropping the last symbol each take constant time.
\end{lemma}
\begin{proof}
\emph{Compare.} Compare the pairs $(\pre(v),b)$ lexicographically. Different old
prefixes are ordered correctly by Lemmas~\ref{lem:fresh-minimum}
and~\ref{lem:prefix-order}. Equal old prefixes have equal $b$ and, as the
lengths agree, equal fresh parts.

\emph{Push.} Appending $i$ increases $D$ and keeps the handle.

\emph{Pop.} A pop is applied only when $D>0$. Every handle with $b>0$ satisfies
$\tdepth(\parentof(v))<b\le\tdepth(v)$, and a handle with $b=0$ uses the root.
If $b<D$, the pop shortens the fresh part and the handle stays. If $b=D$, let
$b':=b-1$ and $u:=\parentof(v)$. The new handle is $(v,b')$ if $b'>\tdepth(u)$,
and $(u,b')$ otherwise, in which case the new prefix is exactly $\lambda(u)$.
Both cases preserve the invariant.

The return at the root creates the key $(i)$, with handle $(\mathrm{root},0)$.
Failure has a separate value, which all operations preserve.
\end{proof}

For example, layer $1$ of Table~\ref{tab:overview-trace} starts with the trie of
Figure~\ref{fig:overview-operations}a, whose preorder is $()$, $3$, $322$,
$3222$, $33$, $332$. The value $P_Q(1)=322$ has the handle $(322,3)$ at length
$3$. Its pop $32$ has the handle $(322,2)$, since $2>\tdepth(3)=1$. The other
candidate for $C_{a_1}(1)$ is $C_F(1)=31$, the old prefix $3$ followed by one
fresh symbol, with the handle $(3,1)$. As $\pre(3)<\pre(322)$, the handles order
$31<32$, and the push of $1$ gives $C_{a_1}(1)=321$ with the handle $(322,2)$ at
length~$3$.

\paragraph{Preparing the next layer.}
Once $A$ holds the values $P_e(i-1)$, the symbol $i$ becomes old. Before the
next layer, every key in $A$ needs its own vertex, so that its handle can start
the layer with $b=D$.

\begin{lemma}[Rebuilding the trie]\label{lem:frozen-rebuild}
From the frozen trie of layer $i$ and the handles of the values $P_e(i-1)$, the
compressed trie of these values, its preorder numbers, and a handle
$(u,\tdepth(u))$ for every value can be built in $O(m)$ time and words.
\end{lemma}

The construction sorts the handles once and adds two kinds of vertices: the old
prefixes that the handles use, exposed inside their edges, and one chain of
fresh symbols below each such prefix. Pruning and renumbering then give the new
trie. Appendix~\ref{app:frozen} gives the details and an example.

\begin{proposition}[The scan]\label{thm:evaluator}\label{compact:theorem}
For a fixed $J$, the scan finds the interval of $A_J(s)$ for a given start $s$,
or for the least start $s$ whose answer is not a failure, in $O(m(n+1))$ time
and $O(m)$ auxiliary words. Storing the chosen action of every evaluated state
costs $O(m(n+1))$ additional bits.
\end{proposition}
\begin{proof}
The nullability bits and templates take $O(m)$ time and words by
Lemma~\ref{lem:empty-template}. Each layer then costs $O(m)$ by
Lemmas~\ref{alg:schedule}, \ref{lem:handles}, and~\ref{lem:frozen-rebuild}, and
the arrays take $O(m)$ words. The end position of a successful root entry is the
first coordinate of its key, which its slot carries beside the handle.
Inspecting the starts in decreasing order needs one saved interval. There are
$O(m(n+1))$ evaluated states, and each chooses among at most two successors,
so all choices fit in $O(m(n+1))$ bits.
\end{proof}

\section{Reconstruction in linear space}\label{sec:outputs}

Let $[s,t)$ be the interval of the answer found by
Proposition~\ref{thm:evaluator}, let $T$ be its preferred parse, and set
$\ell:=t-s$. Each step of $T$ follows from a choice made during the scan. The
difficulty is to recover these choices after the scan has discarded the layers
that made them.

\subsection{Replaying stored choices}\label{sec:queries}

One solution keeps the choices. For a common set $J$ of permitted end positions,
record one bit for the choice at each fork, together with the outcome at each
start: failure, an empty match, or a positive match. From any successful start,
follow the chosen actions and emit $\nu_i(x)$ at each inserted empty parse. This
produces the tree in depth-first order, in time proportional to its size. These
\emph{choice bits} take $O(m(n+1))$ bits, and the replay uses $O(m)$ additional
words. Replaying several starts costs the sum of their output sizes.

For a single answer we do not need the choice bits: the scan keeps only
$[s,t)$, which takes $O(m)$ words even on failure. If $s=t$, the selected parse
is $\nu_s(E)$. Otherwise we solve full matching for $E$ on the slice $w[s:t)$,
which has the same preferred parse. The rest of this section does so in
$O(m+\ell)$ words, recording the visits to one node at a time.

\subsection{Observing one node}

For a node $F$ of $E$, let $\spans_F(T)$ be the list of the intervals of the
positive visits to $F$ in $T$, in depth-first order. Visits to one node cannot
nest, so these intervals are disjoint and occur from left to right. For the body
$F$ of $(aa?)^*$ on \texttt{aaa} (Figure~\ref{fig:overview-operations}b), the
list is $([0,2),[2,3))$.

\begin{lemma}[Positive visits to one node]\label{out:projection}
For a given node $F$, the scan can also return $\spans_F(T)$ on success, using
$O(n+1)$ additional words and $O(m(n+1))$ total time.
\end{lemma}
\begin{proof}
Each slot also stores a pointer to the list of events at $F$ along its preferred
continuation, copied from the chosen successor. At $P_F(i)$ and at $C_F(i)$, the
scan prepends an opening or a closing event at $i$, respectively. The pointer
belongs to the slot rather than to the key, since distinct continuations can
have equal keys. Induction in evaluation order gives the events in depth-first
order. Only these two states allocate records at each position, and every list
shares its tail with the list of its successor. When the scan keeps a root
interval, it also keeps its pointer. Consecutive opening and closing events
delimit the requested intervals, and an empty root gives the empty list.
\end{proof}

\subsection{Contracting given intervals}\label{out:contraction-section}

We adapt the split-and-contract strategy of Bille and
G{\o}rtz~\cite{bille2019,bille2022}: we split the expression at a subtree and
take the intervals to contract from the POSIX parse. Fix a node $F$ of $E$ and a
full-matching query for $E$ on a word $w$. Given a list $\Gamma$ of disjoint
positive intervals in increasing order, write $w/\Gamma$ for the word obtained
by replacing each listed substring by one token $\#$, a symbol that is fresh for
this contraction. Let $\varphi$ map each position of $w/\Gamma$ to the
corresponding position of $w$. The expression $E/F$ replaces the subtree at $F$ by a new leaf
$\#_F$. The leaf $\#_F$ has the nullability bit of $F$, and its only positive
action, $P_{\#_F}(q)\to C_{\#_F}(q+1)$, is legal exactly when the next token is
$\#$. An empty visit at $q$, when allowed, expands to $\nu_{\varphi(q)}(F)$.
No other leaf reads $\#$.
In Figure~\ref{fig:overview-operations}b, $F$ is the body of
$(aa?)^*$, $\Gamma=([0,2),[2,3))$, $w/\Gamma=\#\#$, and $E/F=\#_F^*$.

\begin{lemma}[Order-preserving contraction]\label{out:contraction}
Suppose that every interval in $\Gamma$ is feasible for $F$. Expansion is an
order isomorphism from the full parses of $E/F$ on $w/\Gamma$ to the full parses
$T$ of $E$ on $w$ with $\spans_F(T)=\Gamma$ in which every visit to $F$ is the
preferred parse of $F$ on its interval.
\end{lemma}
\begin{proof}
Every marker token must be read by $\#_F$, and every positive visit to $\#_F$
reads one marker. Expansion replaces each such visit by the preferred parse of
$F$ on the corresponding substring of $w$, and each empty visit at $q$ by
$\nu_{\varphi(q)}(F)$. Contraction reverses these
replacements. Both maps preserve positivity and hence canonicality.

The map $\varphi$ is strictly increasing. When two corresponding visits are compared,
their starts agree because their preceding siblings agree. Their lengths are
therefore ordered by their ends, and $\varphi$ preserves that order. Branch tags
and the order of children outside $F$ are unchanged. If the comparison reaches
two visits to $\#_F$ on the same interval, both expand to the same preferred
subtree. Hence the first difference gives the same preference in both trees.
\end{proof}

For $\Gamma=\spans_F(T)$ taken from the selected parse, every visit to $F$ is
already the preferred parse on its interval, by Lemma~\ref{sem:replacement}.
Thus solving full matching for $E/F$ on $w/\Gamma$ recovers the part of the
selected parse outside $F$.

\subsection{The recursive reconstruction}\label{out:reconstruction-section}

In a recursive call, write $E,w$ for the current expression and input token
sequence, and let $T$ be their preferred parse in local coordinates. Set
$M:=|E|$ and $N:=|w|$; a token may stand for several original characters.
Let $\psi:\{0,\ldots,N\}\to\{0,\ldots,n\}$ map local token
boundaries to positions in the original text. This map is strictly increasing;
initially, on the selected slice $[s,t)$, set $\psi(q):=s+q$. Tests and
comparisons use local positions, whereas a newly emitted visit on a local
interval $[i,j)$ receives the original interval $[\psi(i),\psi(j))$. Each token
stores its original endpoints, and an empty input stores its sole endpoint, so
a map lookup takes constant time and the representation uses $O(N+1)$ words.
We measure a call by
its \emph{potential} $MN$.

If $M$ is below a fixed constant or $N\le1$, replay the choice bits as in
Section~\ref{sec:queries}; the $O(M(N+1))$ bits fit in $O(M+N)$ words. Otherwise
choose a proper subtree $F$ with $M/3-O(1)\le|F|\le2M/3+O(1)$: descend from the
root into the largest child until the subtree has fewer than $2M/3$ nodes. Since
every node has at most two children, this subtree has at least $M/3-O(1)$
nodes. Then $|E/F|=M-|F|+1$. We reconstruct the selected parse as follows.
\begin{enumerate}[leftmargin=*,nosep]
\item Run the scan of Lemma~\ref{out:projection} to find $\Gamma=\spans_F(T)$.
\item For each $[a,b)\in\Gamma$, solve full matching for $F$ on $w[a:b)$ with
      boundary map $\psi_{\mathrm{inner}}(q):=\psi(a+q)$. Store a pointer to
      the resulting parse in its marker token, whose original interval is
      $[\psi(a),\psi(b))$.
\item Solve full matching for $E/F$ on $w/\Gamma$ with boundary map
      $\psi_{\mathrm{outer}}:=\psi\circ\varphi$. A positive visit to $\#_F$
      returns its stored parse unchanged, and an empty visit at $q$ emits
      $\nu_{\psi(\varphi(q))}(F)$.
\end{enumerate}
Copies of a node or a marker keep their original identity. Each contraction uses
a fresh marker, so earlier marker tokens keep their meaning, and the original
nullability bits still suffice to emit every template. Every positive marker is
read once in the outer parse, so a completed inner parse already belongs to the
final output and is counted as output space. Lemma~\ref{out:contraction} proves
correctness by induction on the recursion.

\begin{example}[Original positions after contraction]\label{ex:contraction-coordinates}
Contracting the subtree $H:=aa$ in $(aa)(\capture{g}{\eps}\,b)$ on
\texttt{aab} gives the input $\#b$. Its local boundaries $0,1,2$ correspond
to original positions $0,2,3$, respectively. The empty visit to $g$ is at local
position $1$, but its reported span is $[2,2)$. A later contraction composes
its local boundary map with this map; even if the new local map is the identity,
the original position of this empty visit remains $2$.
\end{example}

\begin{lemma}[Cost of the reconstruction]\label{out:linear-space}
Full matching on the selected slice takes $O(m(\ell+1))$ time and $O(m+\ell)$
auxiliary words. Its parse has at most $m\max\{1,\ell\}$ visits.
\end{lemma}
\begin{proof}
\emph{Time.} Preparing a split costs $O(M(N+1))$. Inner intervals of a single
token are solved at once, in total time $O(|F|N)$, and each becomes one marker.
Let $\Lambda$ be the total length of the other inner intervals and $k$ their
number, so that $k\le\Lambda/2$ and the outer word has $N-\Lambda+k$ tokens.
The potentials of the recursive calls sum to at most
\[
 |E/F|(N-\Lambda+k)+|F|\Lambda\ \le\ N\max\{|E/F|,\ |F|+|E/F|/2\}
 \ \le\ (5M/6+O(1))N ,
\]
by both bounds on $|F|$ and by $|E/F|=M-|F|+1$. With a large enough threshold
for the base case, the potential decreases by a constant factor from one level
of the recursion to the next, so the split costs sum to $O(m(\ell+1))$, starting
from the potential $m\ell$. A nonempty base case costs $O(MN)$ before emitting
the original visits, and an empty query emits its template directly.

\emph{Space.} Release the scan and the current buffers after preparing the
inputs of the children. The sizes of the expressions on the recursion stack
decrease geometrically, and all inner intervals of a split share one copy of
$F$. To bound the pending inputs, consider a descent to a child with $N'$
tokens. For the outer child, $N'=N-\Lambda+k$, and the remaining inputs have
total length at most $\Lambda\le2(N-N')$. For an inner child, the total is at
most $N-N'+k$; since $k-1\le(\Lambda-N')/2$, this is at most $3(N-N')/2+1$. The
differences $N-N'$ telescope down the stack from $\ell$, and the additive
constants sum to $O(\log(m+1))$. Pending intervals have positive length, so
their records of constant size fit the same bound. The current call uses
$O(M+N)$ words even while it prepares the inputs of its children. In total this
gives $O(m+\ell)$ auxiliary words.

\emph{Output size.} Give each visit on $[i,j)$ the mass $\max\{1,j-i\}$. For each
edge of $E$, the visits to the child under one visit to the parent have total
mass at most the mass of that parent visit: the positive iterations of a
repetition partition its interval, an empty repetition has at most one
iteration, and every other node has at most one visit to each child. Induction
down $E$ gives at most $\max\{1,\ell\}$ visits to each node. Emitting an
original visit takes constant time, and splicing a marker is charged to its
recursive call.
\end{proof}

\subsection{A compact encoding of the parse}\label{out:captures}

An explicit parse may spend most of its space on visits whose intervals and
children are already determined by $E$. Define $S(T)$ by deleting all empty
subtrees of $T$ and suppressing every vertex with exactly one remaining child.
Each remaining vertex keeps the order of its children and the node of $E$ that it
visits. For an empty match, $S(T)$ is a distinguished empty value. In both cases
we store the pair $(s,S(T))$. The compact tree is represented by records with
a constant number of writable fields per vertex.

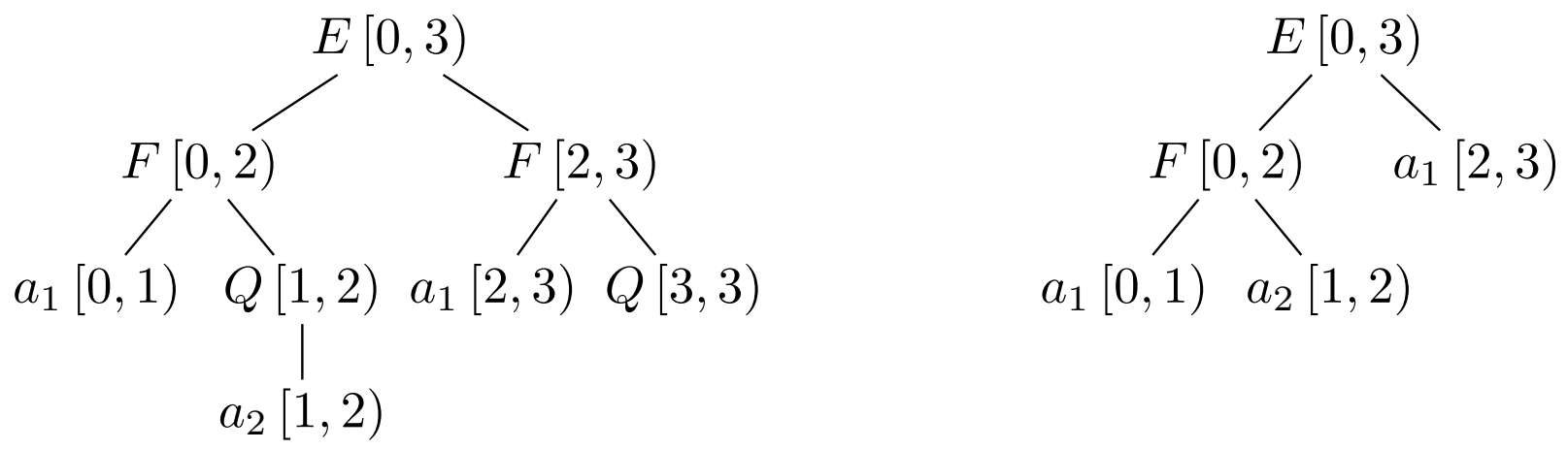
\begin{figure}[H]
\centering
\begin{tikzpicture}[font=\small,x=1cm,y=1cm,
  v/.style={inner sep=1.5pt},
  wire/.style={line width=.45pt}]
  \node[v] (e) at (2.2,0) {$E\,[0,3)$};
  \node[v] (f1) at (0.9,-.85) {$F\,[0,2)$};
  \node[v] (f2) at (3.5,-.85) {$F\,[2,3)$};
  \node[v] (a1) at (0.2,-1.7) {$a_1\,[0,1)$};
  \node[v] (q1) at (1.6,-1.7) {$Q\,[1,2)$};
  \node[v] (a2) at (1.6,-2.55) {$a_2\,[1,2)$};
  \node[v] (a1b) at (2.9,-1.7) {$a_1\,[2,3)$};
  \node[v] (q2) at (4.2,-1.7) {$Q\,[3,3)$};
  \draw[wire] (e)--(f1) (e)--(f2) (f1)--(a1) (f1)--(q1) (q1)--(a2)
              (f2)--(a1b) (f2)--(q2);
  \node at (2.2,-3.3) {(a) The parse $T$};
  \begin{scope}[xshift=7.2cm]
    \node[v] (E) at (1.5,0) {$E\,[0,3)$};
    \node[v] (F) at (0.7,-.85) {$F\,[0,2)$};
    \node[v] (B) at (2.4,-.85) {$a_1\,[2,3)$};
    \node[v] (A) at (0.0,-1.7) {$a_1\,[0,1)$};
    \node[v] (C) at (1.4,-1.7) {$a_2\,[1,2)$};
    \draw[wire] (E)--(F) (E)--(B) (F)--(A) (F)--(C);
    \node at (1.2,-3.3) {(b) The compact tree $S(T)$};
  \end{scope}
\end{tikzpicture}
\caption{The preferred parse of $(aa?)^*$ on \texttt{aaa}, with eight visits,
and its compact tree, with $2\ell-1=5$ vertices. The compact tree deletes the
empty option and suppresses the visits with one remaining child; the intervals
in (b) are implied by the leaves.}
\label{fig:compact}
\end{figure}

\begin{lemma}[Size and decoding]\label{out:skeleton}
Let $T$ be a preferred parse on $[s,t)$, with $\ell=t-s$. If $\ell>0$, $S(T)$
has at most $2\ell-1$ vertices. Given $E$ and $(s,S(T))$, a decoder recovers $T$
in depth-first order in $O(|T|)$ time after $O(m+\ell)$ preparation, using
$O(m)$ additional words.
\end{lemma}
\begin{proof}
Assume first that $\ell>0$. Every leaf of $S(T)$ is a visit to a literal and reads one character, so there
are $\ell$ leaves. Every other vertex has at least two children, so there are at
most $\ell-1$ of them. Counting the leaves below each vertex in postorder gives
every vertex its interval, starting at $s$. These counts and intervals are
stored in the writable fields of the compact tree.

To decode a positive visit to a node $e$, let $x$ be the node of $E$ visited by
the supplied vertex of $S(T)$. If $e=x$, emit the leaf or the branching visit;
the latter is a concatenation with two positive children or a repetition with
at least two positive iterations, so the nodes and intervals of its children
are determined. If $e\ne x$, exactly one child $f$ of $e$ in $E$ has $x$ in its
subtree, and a preorder numbering of $E$ identifies $f$ in constant time. Decode
$f$ with the same vertex of $S(T)$. At an alternation, $f$ gives the branch; at
a repetition, $f$ is the body of a single iteration; and a capture has the
interval of its child. At a concatenation, the other child $y$ is empty and is
emitted as $\nu_i(y)$ at the appropriate position $i$. For an empty match, emit
$\nu_s(E)$. By Lemma~\ref{lem:empty-template}, every deleted empty subtree is
the template emitted by these rules, so they reverse every deletion and
suppression. Each frame keeps a pointer to its next child, so the traversal
uses at most the depth of $E$ in stack words.
\end{proof}

For example, when the decoder of Figure~\ref{fig:compact} reaches the second
iteration of $F$, the supplied vertex visits $a_1$, so the decoder enters $a_1$
and emits $\nu_3(Q)$ for the empty option. As another example, $(a^*)^*$ on
\texttt{aa} admits one or two positive outer iterations. Both literal leaves
visit the same node, but the branching vertex visits the inner star in the
selected parse and the outer star in the other, so the two compact trees differ.

The compact tree also has a short binary encoding. Number the nodes of $E$ in
preorder, and write $S(T)$ with balanced parentheses, which take two bits per
vertex, and one node number of $\lceil\log_2(m+1)\rceil$ bits per vertex. Prefix
a self-delimiting encoding of $s+1$ and a bit that marks an empty match. With $E$
shared, $(s,S(T))$ then takes $O(\log(n+2)+\ell\log(m+1))$ bits, and decoding
needs only $E$ and these bits. Decoding first constructs the $O(\ell+1)$-word
record representation and then applies Lemma~\ref{out:skeleton}.

\paragraph{Building the compact tree during the recursion.}
In a base case, a positive literal produces a leaf and an empty subtree produces
an empty result. A visit with exactly one positive child returns the pointer of
that child. A visit with two or more returns a fresh vertex labelled with its
original node and the ordered list of those pointers; a repetition appends the
pointer of each iteration once. At a contraction, the compact result of each
inner problem is put into its marker token. A positive marker returns that
pointer directly, with no new leaf, and an empty marker yields the empty result.
Replacing a positive marker leaf by a nonempty tree preserves the number and the
order of the positive children at every outer ancestor. These rules therefore
commute with marker expansion and produce $S(T)$, by induction on the
recursion. Every allocated vertex survives in the final answer, including those
held by pending inner solutions, so all vertices and their lists of children
occupy $O(\ell)$ words in total. The costs of Lemma~\ref{out:linear-space} still
apply.

\subsection{The capture vector}\label{out:capture-vector}

By Lemma~\ref{out:last-visit}, one record per capture suffices to read the
capture vector from the visits of $T$ in depth-first order. Number the visits in
this order, starting from $1$. For each capture $g$, keep the number $\tau_g$
and the interval of its last visit, and set $\tau_0:=0$ for the synthetic root.
Then, in parent-before-child order and with $p=\parcap(g)$, set $\capv_T(g)$ to
the interval of the last visit to $g$ if $\capv_T(p)\ne\bot$, $\tau_g$ exists,
and $\tau_g>\tau_p$; otherwise set $\capv_T(g)=\bot$. Once the last visit to $p$
has started, every later visit to $g$ lies inside it, so this test implements
Lemma~\ref{out:last-visit}.

\begin{proposition}[Single-match output]\label{out:refined}
For the queries of Theorem~\ref{thm:main}, the compact parse and the capture
vector, together or separately, take $O(m(n+1))$ time, including preprocessing.
The algorithm uses $O(m+\ell)$ auxiliary words on success and $O(m)$ on
failure. Expanding a successful parse adds $O(|T|)$ time.
\end{proposition}
\begin{proof}
Run the scan of Proposition~\ref{thm:evaluator} on the whole text, and then the
reconstruction of Lemma~\ref{out:linear-space} on the selected slice, building
the compact tree as above. This takes $O(m(n+1))$ time and $O(m+\ell)$ auxiliary
words. For the full parse, decode $(s,S(T))$ by Lemma~\ref{out:skeleton}. For
the capture vector, decode $T$ and visit it in depth-first order without
storing it, keeping the records above. This takes $O(|T|)=O(m\max\{1,\ell\})$
time by Lemma~\ref{out:linear-space}, and $O(m)$ additional words.
\end{proof}

\begin{proof}[Proof of Theorem~\ref{thm:main}]
Apply Proposition~\ref{out:refined} with $\ell\le n$ and $|T|=O(m(\ell+1))$, by
Lemma~\ref{out:linear-space}.
\end{proof}

\begin{example}[An output lower bound with a unique parse]\label{ex:output-lower-bound}
Let $U_0:=a$, $U_{j+1}:=U_j\cdot b?$, and $E_k:=U_k^*$, which has $3k+2$ nodes.
On $a^n$ with $n\ge2$, each iteration must read one $a$, and every option is
empty. The unique parse has $1+n(2k+1)$ visits, whereas its compact tree
consists of the outer star and its $n$ literal leaves. Writing the explicit
parse can thus take $\Theta(mn)$ time, even for an expression without captures
and an input with a single parse.
\end{example}

\section{Reporting matches from several starts}\label{sec:further}

Consider $(\capture{g}{a})^*$ on the text \texttt{aab}. The first two starts
give different whole matches, but both enter the same capture at position $1$,
read the second \texttt{a}, and report the same last capture. The last two
starts have an empty whole match and no visit to $g$.
\begin{center}
\small
\begin{tabular}{@{}lcccc@{}}
\toprule
Start & $0$ & $1$ & $2$ & $3$\\
\midrule
Selected interval & $[0,2)$ & $[1,2)$ & $[2,2)$ & $[3,3)$\\
Reported $g$ & $[1,2)$ & $[1,2)$ & $\bot$ & $\bot$\\
\bottomrule
\end{tabular}
\end{center}
A nonoverlapping search skips start $1$ after reporting $[0,2)$, whereas an
all-start query reports all four columns. The two kinds of output call for
different algorithms. Throughout this section, $c$ denotes the number of named
captures, excluding the synthetic one.

\paragraph{Nonoverlapping matches.}
After each leftmost-longest match, the search resumes at the end of the match
if it is positive, and one character past its end if it is empty. An empty match
at $n$ ends the search.

\begin{corollary}[Nonoverlapping matches]\label{cor:nonoverlap}
All nonoverlapping matches, with their parses or capture vectors, take
$O(m(n+1))$ total time and $O(m+n)$ auxiliary words. With $E$ shared, their
compact parses together occupy $O(n+1)$ words of output.
\end{corollary}
\begin{proof}
One scan stores the end position of the answer, or failure, at every start.
Visit the starts in increasing order, applying the resumption rule, and
reconstruct each selected interval by full matching on its slice. There are at
most $n+1$ intervals, and their lengths sum to at most $n$, so summing
$O(m(\ell+1))$ over the selected lengths $\ell$ gives the time bound. Reusing
the space of the reconstruction gives the space bound. Each positive compact
tree has at most $2\ell-1$ vertices, and an empty match and a match header each
take constant space. Returning the capture vector of a match adds $O(c+1)$ time,
which is $O(m(n+1))$ in total.
\end{proof}

\paragraph{Overlapping starts.}
Fix a common set $J$ of permitted end positions with constant-time membership
tests. For every start $i$ we now report the interval of $A_J(i)$ and the
capture vector of its parse, or failure. Let $d$ be the largest number of literal
nodes of $E$ with the same letter, or zero if $E$ has no literals.

\begin{proposition}[Capture vectors at every start]\label{thm:all-start-captures}
The intervals and capture vectors of all answers $A_J(i)$, $0\le i\le n$, can be
reported in $O(m(n+1))$ total time and $O(m+dc)$ auxiliary words, including
preprocessing. The answers may be streamed in decreasing order of start or
written to slots indexed by start.
\end{proposition}

For $a^*$ on $a^n$, writing a separate compact tree for every start takes
$n(n+1)/2$ leaf writes; the choice bits of Section~\ref{sec:queries} represent
all these trees together. For capture vectors, two starts whose paths reach the
same state share all later events. In the example above, starts $0$ and $1$
share the last visit to $g$. The chosen successors therefore form a tree
rooted at acceptance, and Appendix~\ref{app:all-start-captures} traverses it once,
block by block, keeping only the events nearest to acceptance. With $c$ nested
captures around $a^*$, the output alone has $(n+1)c$ capture entries.

\section{Related work and scope}\label{sec:related}

Table~\ref{tab:related} compares bounds when both the expression and the text
are inputs. Two algorithms may choose the same matching interval and still
report different captures on that interval. Other recent algorithms compute
the overall leftmost-longest span~\cite{varatalu2025} or return all parses in
a shared forest~\cite{borsotti2025parallel}, whereas Theorem~\ref{thm:main}
selects the preferred parse with respect to the syntax tree as written.

\begin{table}[h]
\centering\small
\begin{tabular}{@{}>{\raggedright\arraybackslash}p{.28\linewidth}>{\raggedright\arraybackslash}p{.28\linewidth}>{\raggedright\arraybackslash}p{.36\linewidth}@{}}
\toprule
Work & Required output & Worst-case bound and preprocessing \\[3pt]
\midrule
Myers et al.~\cite{myers1998reporting} & Longest substring or an optimal tagged path &
$O(mn)$ for the span or path graph, $O(\gamma 4^kmn)$ for general tag selection \\[15pt]
Okui--Suzuki~\cite{okui2013} & Canonical POSIX tree &
$O(nd(d+p))$ online, with precomputed transition comparisons \\[15pt]
\mbox{Borsotti--Trofimovich}~\cite{borsotti2021} & POSIX submatches &
$O(nm^2t)$ including construction \\[8pt]
Frisch--Cardelli~\cite{frischcardelli2004} & Greedy parse &
$O(mn)$ with suffix-feasibility annotations \\[15pt]
Grathwohl et al.~\cite{grathwohl2013} & Leftmost-greedy parse &
$O(mn)$ with a sequential choice log \\[8pt]
Bille--G\o rtz~\cite{bille2019,bille2022} & Character transitions of an accepting path &
Basic algorithm: $O(mn)$ time and $O(m+n)$ space \\[15pt]
Theorem~\ref{thm:main} & POSIX parse on the syntax tree as written, and captures &
$O(m(n+1))$ including preprocessing, $O(m+\ell)$ auxiliary words on success \\[8pt]
\bottomrule
\end{tabular}
\caption{Selection objectives and joint complexity in the sizes of the
expression and the text. Here $k$ counts designated tags, $\gamma$ bounds the
time to compare their length vectors, and $\ell$ is the length of the selected
match.}
\label{tab:related}
\end{table}

\paragraph{Recognition information and greedy parsing.}
Kearns~\cite{kearns1991} reconstructs a parse from the state sets saved
during recognition; his main presentation recognizes forward and recovers a
parse backward. Frisch and Cardelli~\cite[Section~5]{frischcardelli2004}
compute suffix feasibility right to left and then build a greedy parse left
to right, choosing the first feasible branch; both passes take $O(mn)$ time.
Like Kearns, they save the state sets of recognition rather than competing
partial parses, and they exclude empty iterations of repetitions with nullable
bodies. Our candidate convention instead permits one empty iteration when the
whole repetition is empty (Example~\ref{ex:empty-iteration}).

The distinction in the selection rule already appears when no repetition has
a nullable body. In Table~\ref{tab:context-priority}, both branches of $X$ admit a
full match on every input, while POSIX selects $aa$, then $a$, then $aa$.
Suffix feasibility and a fixed branch priority do not determine these
choices. Our algorithm orders the feasible continuations by the end positions
that POSIX compares, and maintains that order in $O(m)$ time per input
position. Grathwohl et al.~\cite{grathwohl2013} also obtain $O(mn)$ greedy
parsing, using a sequential choice log. Table~\ref{tab:three-parses} gives
another distinction between branch priority and POSIX on three letters.

\paragraph{Longest intervals and optimal tags.}
Myers, Oliva, and Guimar\~aes~\cite[Section~2.1]{myers1998reporting} find a
longest matching substring anywhere in the text. On the token example
\texttt{x=y==z} of Section~\ref{sec:examples}, this gives the later \texttt{==},
whereas leftmost-longest search gives the first \texttt{=}. For a fixed
interval, their Section~2.3 constructs a graph of matching paths. It selects a
path by the sum of the lengths of designated tags, or by the lexicographic order
of their lengths. For the full match of $a^*\capture{g}{a^*}$ on \texttt{aa},
POSIX gives the first star both letters and reports $g=[2,2)$, whereas
maximizing the length of the only tag gives $g=[0,2)$.

Let $k$ count the designated tags, and let $\gamma$ bound the time to compare two
vectors of tag lengths. A lexicographic comparison takes $O(k)$ time, while
maintained sums take $O(1)$ time to compare. Figure~1 of Myers et al.\ gives the
general bound in Table~\ref{tab:related}. Their refinement enumerates only
realizable tag states, at an additional preprocessing cost, and unambiguous tags
in their sense take $O(\gamma mn)$ time. To see where many realizable states
arise, take
\[
 R_k := (\capture{1}{a}\mid a)\cdots
         (\capture{k}{a}\mid a),\qquad w:=a^k.
\]
For each subset of the $k$ tags, take the tagged branch exactly at its indices.
The final vertex thus admits $2^k$ sets of completed tags in their enumeration.
Grathwohl et al.~\cite[Section~8]{kleenex2016} describe the 1998 result as
$O(mn)$ POSIX-disambiguated matching. The single-tag example above separates the
two objectives even on a fixed interval.

\paragraph{Algorithms for POSIX selection.}
Okui and Suzuki~\cite{okui2011,okui2013} compare lengths at ordered tree
positions. We use the definitions and theorem numbering of their 2013
technical report. Their position automaton uses precomputed comparisons between
transitions, and their online bound uses the letter multiplicity $d$ of
Section~\ref{sec:further}. Write $p$ for their count of capture positions and
stars.\footnote{Table~\ref{tab:related} gives the bound of their Theorem~5 and
conclusion. The abstract of the technical report gives $O(n(d^2+p))$.}
Borsotti and Trofimovich~\cite{borsotti2021} extend the positional order and
account for the construction of their tagged Thompson automaton. Their $m$
counts the expression after expanding bounded repetition, and $t$ counts capture
groups and their enclosing subexpressions. Their algorithm uses $O(m^2)$ space.
Proposition~\ref{sem:correspondence-scope} relates both tree orders to ours.

Laurikari~\cite{laurikari2000} introduced automata whose transitions record the
positions of tags. Determinizing these automata permits efficient matching after
ahead-of-time or just-in-time compilation~\cite{borsotti2022}. Backward scanning
also appears in Cox's prototype~\cite{cox2009} and in the partial-derivative
matcher of Sulzmann and Lu~\cite[Section~4.4 and Proposition~4.4]{sulzmann2010}.
Their 2010 manuscript explains why backward processing can avoid retaining
partial-parse histories, but its time bound treats the pattern as fixed.
Partial derivatives also underlie their later submatching
algorithm~\cite{sulzmann2012}. Lemma~\ref{lem:fork} and the keys of
Section~\ref{sec:evaluation} give a joint bound in $m$ and $n$. The choice rule remains essential: on
$(\capture{g}{aa\mid a})^*$ and
\texttt{aaaaa}, the preferred iterations have lengths $2,2,1$, giving
$g=[4,5)$, whereas Cox's matching core reports $[3,5)$ in the counterexample of
Borsotti and Trofimovich~\cite[Section~1]{borsotti2021}.

\paragraph{Other conventions for POSIX values.}
Derivative-based parsers reconstruct POSIX lexical values by working backward
from the derivatives~\cite{sulzmann2014,urban2023}. Tan and
Urban~\cite{tanurban2023} verify an algorithm for bit-encoded values and bound
the syntax-tree sizes of simplified derivatives for each fixed expression;
the bitcode annotations may still grow with the input. Empty repetition shows a
difference in the output conventions (Example~\ref{ex:empty-iteration}). On
$\eps$, the selected parse of $(\capture{g}{a?})^*$ has one empty iteration, as
Okui and Suzuki~\cite[Definition~3]{okui2013} allow, and reports $g=[0,0)$.
Urban's positive-iteration convention gives the empty list here, so the value
contains no match of $g$. Berglund, Bester, and van der
Merwe~\cite{berglund2021boost} formalize the Boost interpretation, in which
designated groups determine the comparison.

\paragraph{Algorithms that return a witness.}
Dub\'e and Feeley~\cite{dube2000} recover an accepting parse using augmented
automata. Bille and G\o rtz~\cite{bille2019,bille2022} split the automaton and
contract inner substrings to recover the character transitions of an accepting
path. Their basic algorithm takes $O(mn)$ time and $O(m+n)$
space.\footnote{The basic algorithm is given in Section~4 and Theorem~5 of
the conference version~\cite{bille2019}. Its Theorem~6 also gives
$O(mn/\log n+m+n)$ time and $O(m+n)$ space on the word RAM. Table~\ref{tab:related}
lists the basic algorithm whose contraction we use.} On
$(a^*)^*$ and \texttt{aa}, one outer iteration and two successive outer
iterations use the same literal transition twice. POSIX selects one iteration,
whose first body has length two. Our contraction lemma
(Lemma~\ref{out:contraction}) explains why splitting at the visits to a chosen
subtree preserves that preference.

\paragraph{Interval expressions and back-references.}
POSIX also defines interval expressions $e\{k,l\}$, which match $k$ to $l$
consecutive matches of $e$, and, in basic regular expressions, back-references
$\backslash1,\ldots,\backslash9$, which match the text of an earlier
group~\cite[Sections~9.3.6 and~9.4.6]{posix2024regex}. Expanding an interval
expression into copies of $e$ can make the expression exponentially larger:
$k$ nested squarings $(\cdots(a\{2\})\cdots)\{2\}$ denote $a^{2^k}$. Squaring
needs only the count $2$, well within the limit \texttt{RE\_DUP\_MAX} that
POSIX places on counts. Unless the Orthogonal Vectors Conjecture fails, no
algorithm matches expressions with nested squaring in
$O(n^{2-\varepsilon}\operatorname{poly}(m))$ time for any $\varepsilon>0$, and
hence not in $O(mn)$ time, even without star~\cite[Theorem~1]{nogami2026hardness}.
Our states also rely on a property of star and plus: a repetition can stop
after any positive iteration. Consider an exact twofold repetition of
$a\mid aa$ on \texttt{aa}. A return at position~$2$ can follow one two-letter
iteration or two one-letter iterations, and only the second can finish the
repetition. A direct extension of our states would therefore have to
distinguish the two histories. Counting-set automata avoid the expansion for a
substantial class of interval expressions~\cite{turonova2020counting}.

With back-references, the text captured earlier decides whether a continuation
succeeds, so the preferred continuation no longer depends only on the current
node and position (Lemma~\ref{lem:history-independence}). With an unbounded
number of referenced groups, matching is NP-complete~\cite{aho1990}; POSIX
basic regular expressions allow at most nine.\footnote{For a
fixed number $k$ of capture variables, the standard dynamic program takes
$O(mn^{2k+1})$ time~\cite[Section~2.3]{kumabe2026backreferences}.}
A single group with
\emph{straight-line} back-references, where the group and its references occur
only at the top level of a concatenation, already rules out
$O(n^{2-\varepsilon}\operatorname{poly}(m))$ time under the same
conjecture~\cite[Theorem~6]{nogami2026hardness}.
Kumabe and Uezato study a variable-based model that also permits repeated
assignments and self-referential updates: for example,
$\capture{x}{\backslash x\,\backslash x}$ doubles the previous contents of
$x$. Under the Strong Exponential Time Hypothesis, some fixed expression
with one variable in that model cannot be matched in $O(n^{2-\varepsilon})$
time for any $\varepsilon>0$~\cite[Theorem~5]{kumabe2026backreferences}.
This result does not by itself transfer to a single POSIX group. Restricted
uses can still be fast:
for each fixed expression that executes at most one back-reference in any
computation, matching takes $O(n\log^2n)$
time~\cite[Theorem~8]{kumabe2026backreferences}.

\bibliographystyle{posixalpha}
\bibliography{bib,related,output}

\clearpage
\appendix
\section{Summary of notation}\label{app:notation}

\begin{center}
\small
\renewcommand{\arraystretch}{1.05}
\begin{tabular}{@{}p{.25\linewidth}p{.55\linewidth}p{.14\linewidth}@{}}
\toprule
Notation & Meaning & Defined in\\
\midrule
\multicolumn{3}{@{}l}{\emph{Expressions, texts, and parses}}\\
$E$, $m$ & the expression and the number of its nodes & \S\ref{sec:semantics}\\
$w$, $n$ & the text and its length & \S\ref{sec:semantics}\\
$\capture{g}{e}$, $0$ & capture named $g$; the synthetic capture of the whole match & \S\ref{sec:semantics}\\
$\RepExpr_{\lo,\hi}(e)$ & star, plus, or option, with counts $(\lo,\hi)$ & \S\ref{sec:semantics}\\
$w[i]$, $[i,j)$, $w[i:j)$ & character after position $i$; interval; substring & \S\ref{sec:semantics}\\
visit, iteration & vertex of a parse; visit to the body of a repetition & \S\ref{sec:semantics}\\
positive & with a nonempty interval & \S\ref{sec:semantics}\\
$\Parse_e^w(i,j)$ & canonical parses of $e$ on $[i,j)$ & Def.~\ref{sem:def-interval-parses}\\
$T\succ_e U$ & $T$ is preferred to $U$ & Def.~\ref{sem:def-recursive-order}\\
$\codeop(T)$ & depth-first code of a parse & Lemma~\ref{lem:parse-code}\\
$\Opt_e^w(i,j)$ & the preferred parse of $e$ on $[i,j)$ & \S\ref{sec:semantics}\\
$\nu_i(e)$ & preferred empty parse of a nullable $e$; its shape is the template & \S\ref{sec:semantics}\\
$J$, $A_J(i)$ & permitted end positions; answer at the start $i$ & Def.~\ref{sem:def-query}\\
$\parcap(g)$ & nearest capture above $g$ in $E$, or $0$ & Def.~\ref{cap:projection}\\
$\rep_T(g)$, $\capv_T(g)$ & reported visit to $g$ and its interval, or $\bot$ & Def.~\ref{cap:projection}\\
$\lastv_T(g)$ & last visit to $g$ in depth-first order & \S\ref{sec:semantics}\\
\midrule
\multicolumn{3}{@{}l}{\emph{Traversal and keys}}\\
$G_w$, $\acc$ & traversal graph; accepting state & \S\ref{sec:continuations}\\
$P_e(i)$, $C_e(i)$ & entry and return states; also the keys of their preferred continuations, or $\bot$ & \S\ref{sec:continuations}\\
history, continuation & path into a state; path from a state to $\acc$ & \S\ref{sec:continuations}\\
$K(\rho)$, $D(e)$ & end positions of the open visits; their number at a state of $e$ & \S\ref{sec:continuations}\\
$W(\rho)$, $\rho\succ_\sigma\rho'$ & remaining code; continuation preference at state $\sigma$, comparing $K$ before $W$ & \S\ref{sec:continuations}\\
fork & state with two possible next actions & \S\ref{sec:continuations}\\
$\push_i$, $\pop$ & append $i$; drop the last coordinate & \S\ref{sec:continuations}\\
$\firstmax$ & first candidate with the maximum key & \S\ref{sec:continuations}\\
$\mathcal I(e)$, $\mathcal R_e(i)$ & candidates at an entry and at a return & \S\ref{sec:continuations}\\
\midrule
\multicolumn{3}{@{}l}{\emph{Evaluation of a layer}}\\
old prefix, fresh symbol & part of a key in layer $i$ before its copies of $i$; the symbol $i$ & \S\ref{sec:key-order}\\
$\lambda(v)$, $\tdepth(v)$ & word from the root of the trie to $v$; its length & \S\ref{sec:key-order}\\
$\pre(v)$, $\parentof(v)$ & preorder number and parent of a trie vertex & \S\ref{sec:key-order}\\
$(v,b)$ & handle of the key $\lambda(v)[0:b)\,i^{D-b}$ & \S\ref{sec:frozen}\\
\midrule
\multicolumn{3}{@{}l}{\emph{Outputs}}\\
$\ell$ & length of the selected match & \S\ref{sec:outputs}\\
$\spans_F(T)$ & intervals of the positive visits to $F$ & \S\ref{sec:outputs}\\
$w/\Gamma$, $E/F$, $\#_F$ & contracted word and expression; marker leaf & \S\ref{out:contraction-section}\\
$\varphi$, $\psi$ & boundary maps to the current input and to the original text, respectively & \S\ref{out:contraction-section}, \S\ref{out:reconstruction-section}\\
$S(T)$ & compact tree of the preferred parse & \S\ref{out:captures}\\
$c$, $d$ & number of named captures; most literals with one letter & \S\ref{sec:further}\\
$\Phi$, $\Psi$ & translations to the trees of Okui--Suzuki and of Borsotti--Trofimovich & App.~\ref{sec:equivalence}\\
$\dist(v)$ & number of transitions from $v$ to acceptance & App.~\ref{app:all-start-captures}\\
\bottomrule
\end{tabular}
\end{center}

\section{Correspondence with positional orders}\label{sec:equivalence}

This appendix proves Proposition~\ref{sem:correspondence-scope}. We first
express the preference of Definition~\ref{sem:def-recursive-order} by lengths
at tree addresses, and then translate parses into the trees of Okui and
Suzuki~\cite{okui2013} and of Borsotti and Trofimovich~\cite{borsotti2021}.
Throughout, fix a node $e$ and an interval $[i,j)$ of $w$, and let
$x:=w[i:j)$. Both conventions ignore captures and have no node $\emptyre$; we
add $\emptyre$ to them as a node without trees.

\subsection{Positional profiles}

\begin{definition}[Addresses and profiles]\label{eq:def-profile}
Let $T$ be a parse, and contract every visit to a capture into its only child.
An \emph{address} is a finite sequence of natural numbers, and the root of $T$
has the empty address. Below a vertex at the address $p$, the child of an
alternation has the address $p0$ for the left branch and $p1$ for the right
branch, the children of a concatenation have the addresses $p0$ and $p1$, and
the iterations of a repetition have the addresses $p0,p1,\ldots$ in order. Let
$\dom(T)$ be the set of addresses of the vertices of $T$, and $T|_p$ the
subtree at $p\in\dom(T)$. The \emph{profile} of $T$ is the function on all
addresses
\[
 \norm_T(p):=
 \begin{cases}
  \len(T|_p), & p\in\dom(T),\\
  -1, & p\notin\dom(T).
 \end{cases}
\]
Addresses are ordered lexicographically, with a proper prefix before its
extensions; on the addresses of a tree this is the preorder.
\end{definition}

For example, the two parses of $a^*a^*$ on \texttt{a} in
Section~\ref{sec:semantics} have the profile values $1,1,1,0,-1$ and
$1,0,-1,1,1$ at the addresses $\eps,0,00,1,10$, and they first differ at the
address $0$. On the empty word, the two parses of $(\capture{g}{a?})^*$ in
Example~\ref{ex:empty-iteration} have the values $-1$ and $0$ at the address
$0$.

\begin{lemma}[Profiles and preference]\label{eq:profile-order}
Let $T\ne U$ be parses of $e$ with a common start. Then their profiles differ,
and $T\succ_e U$ if and only if $\norm_T>\norm_U$ at the first address where
they differ.
\end{lemma}
\begin{proof}
Induct on $e$, following Definition~\ref{sem:def-recursive-order}. If the
lengths differ, the empty address decides, as test~(i) does. Assume equal
lengths. A literal or $\eps$ has at most one parse of a given length, and
contracting a capture leaves the comparison of its child, which is what the
recursive order does. At an alternation with different branches, the address
$0$ holds a nonnegative norm in the parse of the left branch and $-1$ in the
other parse, and it precedes every address that starts with $1$; this is
test~(ii). Otherwise the children have the addresses $0,1,\ldots$, and every
address that starts with $h$ precedes $h+1$. If the first differing children
have the index $h$, the equal preceding children give them a common start, and
by induction the first difference below $h$ decides as the order of the
children does. If one list of children is a proper prefix of the other, the
first difference is at the next child address, which holds $-1$ for the shorter
list and a nonnegative norm for the longer one, so the extension wins, as in
test~(iii). In each case, different parses have different profiles.
\end{proof}

\Needspace{12\baselineskip}
\subsection{Two tree representations}

\begin{definition}[Trees of Okui and Suzuki]\label{eq:def-os}
Suppose that star is the only quantifier. The translation $\Phi$ contracts the
captures of a parse and maps $\eps$ to the leaf $1$, a literal $a$ to $a$, a
concatenation with children $T_0,T_1$ to $\Phi(T_0)\cdot\Phi(T_1)$, an
alternation with child $T'$ to $L(\Phi(T'))$ or $R(\Phi(T'))$ according to its
branch, and a star with iterations $T_1,\ldots,T_r$ to
$I(\Phi(T_1),\ldots,\Phi(T_r))$. Okui and Suzuki number children from one, so
the child of $R(\cdot)$ has the position $2$. The norm of a tree at a position
is the number of letters in the subtree there, or $-1$ if the position is
absent. A tree is \emph{canonical} if every node $I(t_1,\ldots,t_r)$ with
$r\ge2$ has only nonempty children. Of two trees, the one with the larger norm
at the first differing position is \emph{prior}, and their matching selects the
most prior tree.
\end{definition}

For example, the parse of $a^*$ on \texttt{aa} with two iterations becomes
$I(a,a)$, whose children have the positions $1$ and $2$.

\begin{definition}[Trees of Borsotti and Trofimovich]\label{eq:def-bt}
The translation $\Psi$ supports all three quantifiers. It uses a nil tree
$\bot$ and one ordered constructor $\mathsf T$. It contracts captures, maps
$\eps$ and literals to themselves, a concatenation to
$\mathsf T(\Psi(T_0),\Psi(T_1))$, an alternation to $\mathsf T(\Psi(T'),\bot)$
or $\mathsf T(\bot,\Psi(T'))$ according to its branch, a repetition with
iterations $T_1,\ldots,T_r$, $r\ge1$, to $\mathsf T(\Psi(T_1),\ldots,\Psi(T_r))$,
and a repetition without iterations to $\mathsf T(\bot)$. The node of $E$
determines the kind of each $\mathsf T$ node. We number children from zero,
which renumbers the positions of Borsotti and Trofimovich in an
order-preserving way.
\end{definition}

For example, the left branch of $a\mid b$ becomes $\mathsf T(a,\bot)$. On the
empty word, zero iterations of $\eps^*$ become $\mathsf T(\bot)$, and one empty
iteration becomes $\mathsf T(\eps)$.

\begin{definition}[Positional P-order]\label{eq:def-p-order}
Let $\flattenten(t)$ be the word of the leaves of a tree $t$, where $\bot$
contributes the empty word, and let $\dom(t)$ be its set of positions. Define
\[
 \beta_t(p):=
 \begin{cases}
  -1, & p\in\dom(t)\text{ and }t|_p=\bot,\\
  |\flattenten(t|_p)|, & p\in\dom(t)\text{ and }t|_p\ne\bot,\\
  +\infty, & p\notin\dom(t).
 \end{cases}
\]
Write $t\triangleleft_P t'$ if $\beta_t>\beta_{t'}$ at the first position where
they differ. The $\triangleleft_P$-least tree is preferred. This is the
positional P-order of Borsotti and Trofimovich without their numerical
submatch annotations, which the order does not inspect.
\end{definition}

For example, $\mathsf T(\eps)\triangleleft_P\mathsf T(\bot)$ by $0>-1$ at the
position $0$, and $\mathsf T(\eps)\triangleleft_P\mathsf T(\eps,\eps)$ by
$+\infty>0$ at the position $1$. The value $+\infty$ for a missing iteration is
the one difference from profiles: there, a missing iteration has the norm $-1$
and the longer list wins.

\begin{definition}[Relaxed parses]\label{eq:def-unrestricted}
A \emph{relaxed parse} is defined like a canonical parse
(Definition~\ref{sem:def-interval-parses}), except that two or more iterations
may include empty ones. Let $\mathcal A$ be the set of trees $\Psi(T)$ of the
relaxed parses $T$ of $e$ on $[i,j)$.
\end{definition}

These are the trees that Borsotti and Trofimovich compare. For example,
$\mathsf T(\eps,\eps)$ belongs to $\mathcal A$ for $\eps^*$ on the empty word,
although its two empty iterations are not canonical. Since the number of empty
iterations is unbounded, $\mathcal A$ can be infinite.

\begin{lemma}[Replacement in the P-order]\label{eq:p-context}
Let $T'$ be obtained from a relaxed parse $T$ by replacing a subtree $U$ by a
relaxed parse $U'$ of the same node on the same interval. If
$\Psi(U')\triangleleft_P\Psi(U)$, then $\Psi(T')\triangleleft_P\Psi(T)$.
\end{lemma}
\begin{proof}
The lengths at the proper ancestors of $U$ do not change, and neither do the
positions outside the subtree, including the nil children. If $U$ has the
position $p$ in $\Psi(T)$, and $\Psi(U)$ and $\Psi(U')$ first differ at the
relative position $q$, then $\Psi(T)$ and $\Psi(T')$ first differ at $pq$, with
the same values.
\end{proof}

\begin{lemma}[Removing empty iterations]\label{eq:normalize}
For every relaxed parse $T$ of $e$ on $[i,j)$, there is a canonical parse $U$ of
$e$ on $[i,j)$ with $\Psi(U)=\Psi(T)$ or $\Psi(U)\triangleleft_P\Psi(T)$.
\end{lemma}
\begin{proof}
Induct on $e$. Replacing every child of the root by the parse that induction
provides keeps the intervals and, by Lemma~\ref{eq:p-context}, does not worsen
the tree. Afterwards only the root can violate canonicality, as a repetition
with at least two iterations, some of them empty. If some iteration is
positive, delete all empty ones. At least one iteration remains and the count
does not grow, so the counts stay legal because $\lo\le1$. At the position of
the first deleted iteration, the value changes from $0$ to the positive length
of the next remaining iteration, or to $+\infty$ if none follows. If all
iterations are empty, keep only the first; the value at the position $1$
changes from $0$ to $+\infty$. In both cases all earlier positions are
unchanged, so the tree strictly improves.
\end{proof}

\begin{proposition}[Agreement with positional orders]\label{sem:external-agreement}
The following hold for the node $e$ and the interval $[i,j)$.
\begin{enumerate}[label=(\roman*),leftmargin=*]
\item If star is the only quantifier, $\Phi$ is a bijection from
      $\Parse_e^w(i,j)$ to the canonical trees of Okui and Suzuki for $x$ on the
      node $e$ with its captures contracted, and $T\succ_e U$ if and only if
      $\Phi(T)$ is prior to $\Phi(U)$.
\item With native $*$, $+$, and $?$, for all $T,U\in\Parse_e^w(i,j)$,
      $T\succ_e U$ if and only if $\Psi(T)\triangleleft_P\Psi(U)$.
\item If $\Parse_e^w(i,j)\ne\emptyset$, then $\Psi(\Opt_e^w(i,j))$ is the
      $\triangleleft_P$-least tree of $\mathcal A$.
\end{enumerate}
\end{proposition}
\begin{proof}
(i) The node $e$ determines the contracted captures, and the constructors of
$\Phi(T)$ record the rest of $T$, so $\Phi$ is injective. Conversely, the
lengths of the subtrees of a tree for $x$ determine the intervals of a parse,
and both sides impose the same condition on lists with two or more iterations.
The positions of $\Phi(T)$ are the addresses of $T$ with one added to every
component, which preserves the lexicographic order, and the norms agree with
the profile. Lemma~\ref{eq:profile-order} gives the order.

(ii) Compare $\Psi(T)$ and $\Psi(U)$ with the induction of
Lemma~\ref{eq:profile-order}. The values at corresponding positions are the
same lengths, and a nil branch has the value $-1$ that the profile assigns to
the absent branch. A repetition without iterations has a nil child of value $-1$
at the position $0$, where the profile has $-1$ as well. The two orders could
differ only where one nonempty list of iterations is a proper prefix of the other,
since a missing iteration has the value $+\infty$ in the P-order. This cannot
happen on a common interval: the longer list has at least two iterations, all
positive, so its additional iterations would extend beyond the end of the
shorter list.

(iii) Distinct relaxed parses of $e$ on $[i,j)$ have distinct images under
$\Psi$, and these differ in $\beta$ at some position: different branches place
the nil child differently, and different numbers of iterations differ at the
first missing one. Comparison at the first difference is therefore a strict
total order on $\mathcal A$. By Lemma~\ref{eq:normalize}, every tree of
$\mathcal A$ is matched or beaten by the image of a canonical parse, and by~(ii)
the image of $\Opt_e^w(i,j)$ is the least among these finitely many images.
\end{proof}

Canonicality matters in both directions. Without it,
Definition~\ref{sem:def-recursive-order} would prefer two empty iterations of
$\eps^*$ to one, and the parses of $\eps^*$ on the empty word would form an
increasing chain without a maximum. The P-order instead prefers one empty
iteration to two, so it selects the canonical parse even among relaxed ones.

\section{Rebuilding the trie}\label{app:frozen}

This appendix proves Lemma~\ref{lem:frozen-rebuild}. Every vertex of the trie
stores its depth, its parent, its preorder number, and its children in a doubly
linked list, ordered by their first symbols. The symbols themselves are not
stored. A handle $(v,b)$ with $b>0$ satisfies
$\tdepth(\parentof(v))<b\le\tdepth(v)$, and a handle with $b=0$ uses the root
(Lemma~\ref{lem:handles}). Thus $(v,b)$ names the old prefix
$\lambda(v)[0:b)$, which lies on the edge into $v$ unless $b=\tdepth(v)$. A slot
whose value at $P_e(i-1)$ is not $\bot$ holds such a handle and a key of length
$D=D(e)$; we write $k:=D-b$ for the number of fresh symbols at its end. There
are at most $m$ such slots, and the old trie has $O(m)$ vertices.

\begin{proof}[Proof of Lemma~\ref{lem:frozen-rebuild}]
The construction has four steps.

\emph{Sort.} Sort the slots by $(\pre(v),b,k)$ lexicographically, with stable
counting-sort passes on $k$, then $b$, then $\pre(v)$. All three components are
less than the number of vertices or at most $m$, so this takes $O(m)$ time. By
Lemmas~\ref{lem:fresh-minimum} and~\ref{lem:prefix-order}, the order is the
lexicographic order of the keys, and slots with equal keys are adjacent.

\emph{Expose the old prefixes.} The slots with handles at a vertex $v$ form one
run, in increasing order of $b$. For each distinct $b$ in this run with
$\tdepth(\parentof(v))<b<\tdepth(v)$, create a vertex of depth $b$, and
replace the edge from $\parentof(v)$ to $v$ by the path through these vertices,
in increasing depth. The topmost vertex of the path takes the place of $v$
among the children of $\parentof(v)$. The \emph{base} of a slot is the vertex of
its old prefix: the root if $b=0$, the vertex $v$ if $b=\tdepth(v)$, and the
new vertex of depth $b$ otherwise.

\emph{Attach the fresh symbols.} The slots with a common base $x$ form one run,
in increasing order of $k$. Let $k_1<\cdots<k_r$ be the distinct positive
values of $k$ in this run. Create a path $x,y_1,\ldots,y_r$ with
$\tdepth(y_q)=\tdepth(x)+k_q$, so that $\lambda(y_q)=\lambda(x)\,i^{k_q}$,
and insert $y_1$ as the first child of $x$. This order is correct because $i$
is smaller than every old symbol. Mark the vertex of each slot, which is $x$ if
$k=0$ and $y_q$ if $k=k_q$; no other vertex is marked.

\emph{Prune and renumber.} In postorder, delete every unmarked leaf other than
the root, and suppress every unmarked vertex other than the root that has
exactly one child left. Then number the vertices in preorder, and give each
slot the handle $(u,\tdepth(u))$ of its marked vertex $u$.

The first three steps preserve the words of all existing vertices and the order
of the children, so the trie then contains a vertex for every new key. After
pruning, every leaf is marked, and every unmarked vertex other than the root has
at least two children. The result is therefore the compressed trie of the new
keys, and the new handles satisfy the invariant with $b=\tdepth(u)$. Each step
creates at most one vertex per slot and visits each vertex a constant number of
times, so the construction takes $O(m)$ time and words.
\end{proof}

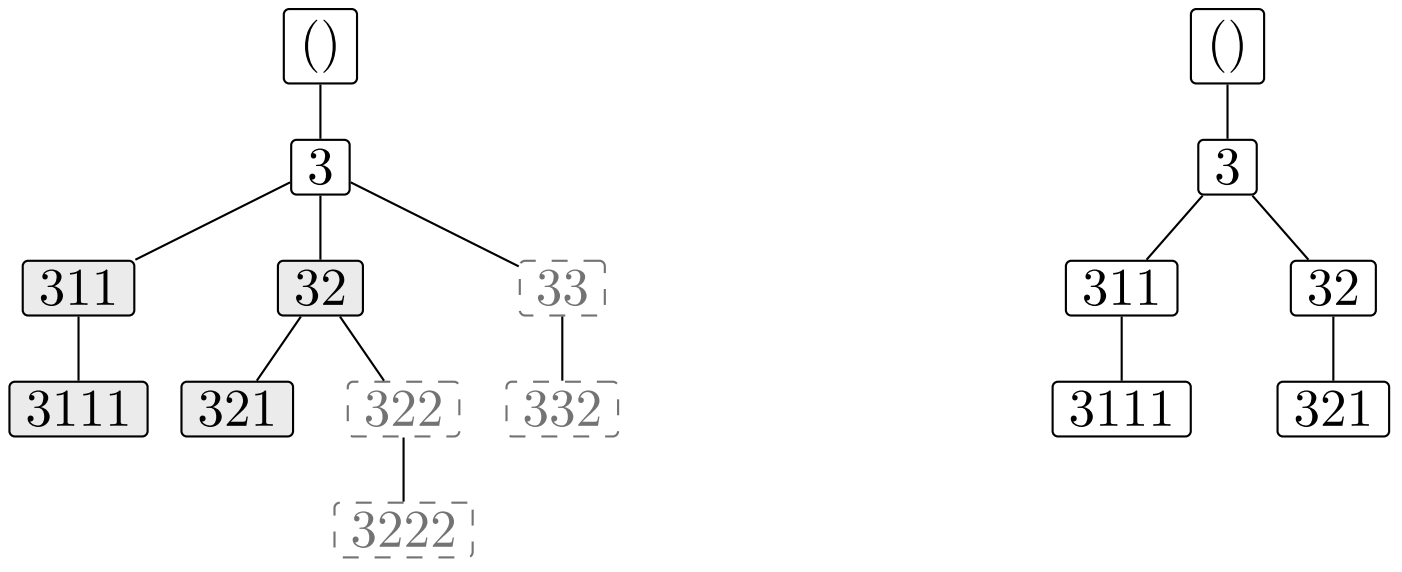
\begin{figure}[ht]
\centering
\begin{tikzpicture}[font=\small,x=1cm,y=1cm,
  key/.style={draw,rounded corners=1pt,inner xsep=3pt,inner ysep=2pt},
  old/.style={key,dashed,text=black!55,draw=black!55},
  new/.style={key,fill=black!8}]
  \node[key] (r) at (0,0) {$()$};
  \node[key] (k3) at (0,-.8) {$3$};
  \node[new] (k311) at (-1.6,-1.6) {$311$};
  \node[new] (k3111) at (-1.6,-2.4) {$3111$};
  \node[new] (k32) at (0,-1.6) {$32$};
  \node[new] (k321) at (-.55,-2.4) {$321$};
  \node[old] (k322) at (.55,-2.4) {$322$};
  \node[old] (k3222) at (.55,-3.2) {$3222$};
  \node[old] (k33) at (1.6,-1.6) {$33$};
  \node[old] (k332) at (1.6,-2.4) {$332$};
  \draw (r)--(k3) (k3)--(k311) (k311)--(k3111) (k3)--(k32) (k32)--(k321)
        (k32)--(k322) (k322)--(k3222) (k3)--(k33) (k33)--(k332);
  \node at (0,-4.0) {(a) After exposing and attaching};
  \begin{scope}[xshift=6cm]
    \node[key] (R) at (0,0) {$()$};
    \node[key] (K3) at (0,-.8) {$3$};
    \node[key] (K311) at (-.7,-1.6) {$311$};
    \node[key] (K3111) at (-.7,-2.4) {$3111$};
    \node[key] (K32) at (.7,-1.6) {$32$};
    \node[key] (K321) at (.7,-2.4) {$321$};
    \draw (R)--(K3) (K3)--(K311) (K311)--(K3111) (K3)--(K32) (K32)--(K321);
    \node at (0,-4.0) {(b) After pruning};
  \end{scope}
\end{tikzpicture}
\caption{Rebuilding the trie after layer $1$ of
Table~\ref{tab:overview-trace}. Shaded vertices are new, and dashed vertices
are old vertices that no key uses any more.}
\label{fig:rebuild}
\end{figure}

For example, the trie of Figure~\ref{fig:overview-operations}a arises in this
way at the end of layer~$2$ of Table~\ref{tab:overview-trace}. The old trie is
the path through $3$, $33$, $333$, and $3333$, and the handles of the five
slots lie at its vertices $3$ and $33$, so no prefix needs to be exposed. The
base $3$ receives the path to $322$ and $3222$, the base $33$ receives $332$,
and pruning removes $3333$ and $333$.

The end of layer~$1$ exercises all four steps. The old trie is that of
Figure~\ref{fig:overview-operations}a, and the slots of $E,F,a_1,Q,a_2$ hold
the keys $3,32,321,311,3111$ with the handles
\[
 (3,1),\quad(322,2),\quad(322,2),\quad(3,1),\quad(3,1)
\]
and the numbers of fresh symbols $0,0,1,2,3$. Sorting gives the runs
$E,Q,a_2$ at the vertex $3$ and $F,a_1$ at the vertex $322$. The second run
exposes the prefix $32$ inside the edge from $3$ to $322$. The base $3$
receives the path to $311$ and $3111$ as its new first child, and the base
$32$ receives $321$. Pruning removes $3222$, $322$, $332$, and $33$
(Figure~\ref{fig:rebuild}). The new preorder is $()$, $3$, $311$, $3111$, $32$,
$321$, and every new handle has $b=D$.

\section{Capture vectors at every start}\label{app:all-start-captures}

This appendix proves Proposition~\ref{thm:all-start-captures}. We first describe a
single traversal that reads every capture vector, and then bound its space by
splitting the positions into blocks. Throughout this appendix, entry and return
states range over those evaluated by Figure~\ref{fig:fused-visit}:
$P_e(i)$ for $0\le i\le n$ and $C_e(i)$ for $1\le i\le n$.

\paragraph{The tree of chosen successors.}
Call a state \emph{successful} if its value is not $\bot$. The chosen action of
a successful state leads to a successful state, or from $C_E(i)$ to $\acc$; call
this target its \emph{successor}. Every successful state therefore reaches
$\acc$, and the successful states form a tree rooted at $\acc$ in which the
parent of a state is its successor. Let $\dist(v)$ be the number of
transitions from $v$ to $\acc$. By Proposition~\ref{alg:correct}, if
$P_E(s)\ne\bot$, the path from $P_E(s)$ to $\acc$ traverses the parse of
$A_J(s)$.

The \emph{label} of a successful state $v$ lists the capture events of $v$ and
of its chosen transition in depth-first order: $P_g(i)$ opens a capture $g$ at
$i$, $C_g(i)$ closes it at $i$, and a transition that inserts $\nu_i(x)$
\emph{invokes} this empty parse, which opens and closes each of its captures at
$i$. A label has at most two entries. The $j$-th entry of the label of $v$
receives the \emph{stamp} $(-\dist(v),j)$, and stamps are compared
lexicographically. Along a path, the stamps increase, and the events of one
invocation share its stamp.

\paragraph{One traversal.}
Traverse the tree depth-first from $\acc$, that is, backward along the paths.
For every capture $g$, keep a record $\mathrm{open}[g]$ and a record
$\mathrm{close}[g]$. Each is empty or holds a stamp, a position, and the state
that wrote it. On entering a state $v$, write its events, from the last to the
first, into the records that are empty. On leaving $v$, empty the records that
$v$ wrote.

\begin{lemma}[Last events]\label{lem:last-events}
When the traversal is at a state $v$, the records of each capture hold its last
opening and its last closing on the path from $v$ to $\acc$, and are empty if
there is none.
\end{lemma}
\begin{proof}
The path from $v$ is $v$ followed by the path from its successor, and the
events of $v$ precede all later events. Induction on the traversal shows that
a record that is empty at the successor receives the last such event of $v$, if
there is one, and that all other records are kept. Leaving $v$ restores the
records of the successor.
\end{proof}

At a successful root entry $P_E(s)$, the records describe the parse $T$ of
$A_J(s)$. Visits to one capture do not nest, so the last opening and the last
closing of $g$ belong to its last visit. In parent-before-child order, with
$p=\parcap(g)$, report the interval of this visit as $\capv_T(g)$ if $g$ has an
opening on the path and either $p=0$, or $\capv_T(p)\ne\bot$ and the last
opening of $g$ has a stamp at least that of the last opening of $p$; otherwise
report $\bot$. This is the test of Section~\ref{out:capture-vector}, with stamps
in place of the numbers $\tau$. Every visit to $g$ lies inside one of the
disjoint visits to $p$, so the last visit to $g$ lies inside the last visit to
$p$ exactly when it opens no earlier; equal stamps arise only inside one
invocation, where the visit to $g$ lies inside the visit to $p$. By
Lemma~\ref{out:last-visit}, the test is correct. If $P_E(s)=\bot$, the answer
is a failure or $\nu_s(E)$. A template visits each node at most once, together
with its ancestors, so every capture of $\nu_s(E)$ reports $[s,s)$ and every
other capture reports $\bot$.

In the example of Section~\ref{sec:further}, $(\capture{g}{a})^*$ on
\texttt{aab}, the paths from $P_E(0)$ and $P_E(1)$ meet at $P_g(1)$, which
opens the iteration on $[1,2)$. The traversal passes $C_g(2)$ and $P_g(1)$
before it reaches either start, so it records the closing at $2$ and the
opening at $1$. These records are not empty when it later passes $C_g(1)$ and
$P_g(0)$ on its way to $P_E(0)$, so the earlier events are not written, and
both starts read $g=[1,2)$.

\paragraph{Invocations in constant time.}
An invocation may contain $\Theta(c)$ captures. For each nullable node $x$,
call the children of $x$ that $\nu_i(x)$ visits its \emph{template children}:
both children of a concatenation, the first nullable branch of an alternation,
the child of a capture, and the body of a repetition if the body is nullable
(Lemma~\ref{lem:empty-template}). These links form a forest on the nullable
nodes, and the captures of $\nu_i(x)$ are those in the subtree of $x$. Delete
the subtrees without captures, suppress the noncapture vertices with one child
left, and attach the remaining roots to a new root. The resulting tree $Y$ has
$O(c+1)$ vertices, since each of its leaves is a capture and each other
noncapture vertex, except the new root, has at least two children. A postorder
pass maps each nullable $x$ to the vertex $y(x)$ of $Y$ whose subtree has the
same captures as $\nu_i(x)$, or to nothing if there are none. For example, the
template of $\capture{h}{\eps\capture{u}{\eps}\mid\capture{v}{\eps}}$ takes the
left branch, so one invocation opens and closes $h$ and $u$ but not $v$; in
$Y$, the vertex $u$ is a child of $h$, and $v$ is a child of the new root.

The traversal keeps a record $\mathrm{inv}[y]$ for every vertex $y$ of $Y$,
with the same rules: an invocation of $\nu_i(x)$ is written into
$\mathrm{inv}[y(x)]$ if it is empty and is erased when its writer is left. Thus
$\mathrm{inv}[y]$ holds the last invocation on the path whose captures are
exactly those below $y$. Before a read, one pass over $Y$ from the root gives
each capture $g$ the last invocation recorded at $g$ or above it. The last
visit to $g$ comes from this invocation, with the interval $[i,i)$, or from the
explicit records of $g$, whichever has the later opening. Each label now has
constant size, and each read takes $O(c+1)$ time.

\paragraph{Blocks.}
The tree has $O(m(n+1))$ states, more than the space bound allows. Split the
positions $0,\ldots,n$ into blocks $[L,R)$ of $B$ consecutive positions, the
leftmost block possibly shorter, and assign each state to the block of its
position. The final position $n$ is included, so the rightmost block ends at
$R=n+1$. Positions never decrease along a transition. For $R\le n$, the only
transitions across a block's right boundary are $P_a(R-1)\to C_a(R)$ with
$w[R-1]=a$. Hence its successful states form subtrees that hang from $\acc$
or from at most $d$ such states $C_a(R)$. The rightmost block has no outgoing
boundary transitions, and its subtrees hang only from $\acc$.

The blocks are processed from right to left during the scan. While the scan
computes the values of a block, it stores the successor of each successful
state and, at the root entries, the end of the answer. Once the values at all
positions of $[L,R)$ are known, the traversal of the block starts from $\acc$
with empty records and from every state $C_a(R)$ for which the block to the
right saved a \emph{summary}: its records, including those of $Y$, and its value
of $\dist$.
It restores these records and traverses the states of the block. If $L>0$, it
saves one summary at every successful state $C_a(L)$ with $w[L-1]=a$ for the
next block; the leftmost block saves none. A summary takes $O(c+1)$ words.
The answers of a block are buffered and emitted in decreasing order of start,
or written to their slots directly. The block buffer is auxiliary space;
slots retaining the reported answers are output space.

\begin{proof}[Proof of Proposition~\ref{thm:all-start-captures}]
The tree $Y$, the capture parents, and the templates take $O(m)$ time and
words. Before the scan, compare every literal with every character of $w$ to
find the largest number $d'\le d$ of literals that match one character, in
$O(m(n+1))$ time; take $d'=0$ when $w$ is empty. Set
\[
 B:=\max\{1,\lceil d'c/m\rceil\}.
\]
A block has $O(mB)$ states, so its buffer and its traversal take $O(mB)$ time
and words. It restores and saves at most $2d'$ summaries, which take
$O(d'(c+1))=O(mB)$ time and words, since $d'c\le mB$ and $d'\le m$. Reading a
capture vector takes $O(c+1)$ time per start, $O((n+1)(c+1))=O(m(n+1))$ in
total, and a buffer of $O(B(c+1))=O(m+dc)$ words. If $n+1\ge B$, there are at
most $2(n+1)/B$ blocks, which take $O(m(n+1))$ time in total; otherwise a
single block without summaries takes $O(m(n+1))$ time. The space is
$O(m+mB+d'(c+1))=O(m+dc)$ words. Correctness follows from
Lemma~\ref{lem:last-events}, since the summaries and $\dist$ carry the records
of the states to the right of each block.
\end{proof}

\end{document}